\documentclass[11pt,a4paper]{article}
\pdfoutput=1

\usepackage{jcappub}

\usepackage{graphicx}

\usepackage{float}
\usepackage{dcolumn}
\usepackage{bm}
\usepackage{color}
\usepackage{enumitem}
  \usepackage{tabularx}
  \usepackage{amsmath}
    \usepackage{cases}
   \newcolumntype{C}{>{\centering\arraybackslash}X}
   \newcolumntype{L}{>{\raggedright\arraybackslash}X}
   \newcolumntype{R}{>{\raggedleft\arraybackslash}X}
\setlistdepth{10}

\definecolor{DarkBlue}{rgb}{0,0,0.7} 

\definecolor{DarkRed}{rgb}{0.65,0,0}

\author{Jaume~Garriga and Nikolaos Triantafyllou}
\emailAdd{jaume.garriga@ub.edu, nitriant@icc.ub.edu}
\affiliation{
\fontsize{11pt}{0pt}\selectfont
Departament de F\'isica Qu\`antica i Astrof\'\i sica, Institut de Ci\`encies del Cosmos, Universitat de
Barcelona, Mart\'\i\ i Franqu\`es 1, 08028 Barcelona, Spain
\vspace{1.5mm}
}

\title{
Enhanced cosmological perturbations
and the merger rate of PBH binaries}

\abstract{The rate of merger events observed by LIGO/Virgo can be used in order to probe the fraction $f$ of dark mater in the form of primordial black holes (PBH). 
Here, we consider the merger rate of PBH binaries, accounting for the effect of cosmological perturbations on their initial eccentricity $e$. The torque on the binaries may receive significant contributions from a wide range of scales, that goes from the size of the horizon at the time when the binary forms, down to the co-moving size of the binary. 
Extrapolating the observed plateau in the power spectrum $P_\Phi \approx 10^{-9}$ from cosmological scales down to the co-moving size of binaries, the torque from perturbations is small. In this case,  for $f\gtrsim 10^{-2}$, the distribution of eccentricities is dominated by tidal torques from neighboring PBHs. On the other hand,  in scenarios where PBH are formed from adiabatic perturbations, it is natural to expect an enhancement of $P_\Phi$ at small scales,
where it is poorly constrained observationally. The effect can then be quite significant. For instance, a nearly flat spectrum with amplitude $P_\Phi \gtrsim 10^{-7}$ on scales smaller than $\sim 10 Mpc^{-1}$  gives a contribution $\langle j^2 \rangle \sim 10^3 P_\Phi$, where $j= (1-e^2)^{1/2}$ is the dimensionless angular momentum parameter of the binaries.
This contribution can dominate over tidal torques from neighboring PBHs for any value of $f$.
Current constraints allow for a power spectrum as large as $P_\Phi \sim 10^{-5}$ at the intermediate scales $10^3-10^5 Mpc^{-1}$, comparable to the co-moving size of the binaries at the time of formation. 
In particular, this can relax current bounds on the PBH abundance based on the observed LIGO/Virgo merger rate, allowing for a fraction $f\sim 10\%$ of dark matter in PBH of mass $\sim 30 M_\odot$.
We investigate the differential merger rate $\Delta\Gamma(m_1,m_2)$, as a function of the masses of the binary components, and the corresponding ``universality" 
coefficient \cite{Kocsis} $\alpha = -(m_1+m_2)^2 \partial^2 \ln \Delta\Gamma/\partial m_1\partial m_2$.
For an enhanced power spectrum with spectral index $p$ we find that
$\alpha \approx 30/(32-7p)$ for $0< p \lesssim 2$, and $\alpha \approx 5/3$ for $p\gtrsim 2$. Such values may lie well outside the narrow range $\alpha \approx 1\pm 0.05$ 
characteristic of tidal forces from neighboring PBHs. We conclude that, given a large enough sample of events, merger rates may provide valuable information on the spectrum of primordial cosmological perturbations at currently uncharted lengthscales.}

\begin{document}

\maketitle

\section{Introduction}

The detection of gravitational waves (GWs) from merging black hole binary systems has revived interest in the idea that primordial black holes (PBHs) may be a viable candidate for dark matter (DM). The abundance of PBHs is severely constrained for a wide range of masses (see e.g. \cite{Carr} and references therein), but it could still be significant both for sublunar and stellar masses. In particular there is an active debate on wheter PBHs in the mass range recently detected by LIGO/Virgo collaboration could account for a sizable fraction of DM \cite{Bird,Sasaki1}. 

The observed merger rate \cite{Abbott,Abbott1,Abbott2,Abbott3,Abbott4} $\Gamma\approx 10-100 \ Gpc^{-3} yr^{-1}$ in the range $\sim 5-100 M_{\odot}$ has recently been used in order to place limits on the PBH abundance \cite{Sasaki1,Kam}. In such estimates, it has been assumed that PBH are spatially uncorrelated at the time of formation, and that the dominant contribution to the orbital angular momentum of the binaries originates from tidal forces exerted by other black holes in the neighborhood, around the time when the binary decouples from the Hubble flow \cite{Thorne,Ioka,Kocsis}. With these assumptions, the observational upper bound on the merger rate limits the fraction of PBHs in DM to $f \lesssim 1\%$. Several refinements to this estimate have been considered, including initial spatial correlations of the PBHs \cite{RVV,Guillermo,Riotto,Domcke}, tidal forces from non-relativistic matter perturbations \cite{Eroshenko,Kam,Chen}, as well as the effect of a dark matter dress around the PBHs \cite{KGB}, with similar results for the bound on the PBH abundances. In Ref. \cite{RSVV}, the effect of infalls of neighboring PBHs on the binary has been studied, with the conclusion that this may significantly reduce the observed merger rate.  Also, N-body simulations for the formation and evolution of binaries \cite{RSVV} indicate that for high $f\sim 1$, the rate may be significantly reduced by disruption, through the interaction of binaries with compact N-body systems.

In this paper, we consider the effect of primordial cosmological perturbations on the angular momentum of PBH binaries during the radiation dominated era. We note that, even for a scale invariant spectrum of density perturbations, there is a wide range of scales contributing to the torque. Moreover, the amplitude of the power spectrum $P_\Phi$ is poorly constrained beyond the scale of $10 Mpc^{-1}$, and it could be significantly larger than it is on cosmological scales. In the present context this possibility seems rather natural, since some scenarios for PBH formation\footnote{Not all scenarios for PBH formation require a bump in the power spectrum. For instance, PBH could be created by active seeds such as relic domain walls or false vacuum bubbles produced during inflation \cite{active,active2,active3}, rather than adiabatic perturbations. In such alternative scenarios, an enhancement in the spectrum of cosmological perturbations does not seem to be a necessary feature. The same is true for PBH formation at  post-inflationary phase transitions (see e.g. \cite{Khlopov,Khlopov1,Khlopov2,Oriol} and references therein).} rely on a prominent  enhancement or ``bump" in the power spectrum at relatively short wavelengths, corresponding to the co-moving size of the horizon at the time when PBHs form. In the inflationary context, the height and location of the bump depend on specific features in the inflaton potential. For instance, in one field models, the field may undergo a short period of ultra-slow roll or constant roll as it encounters local extrema on its way down the potential \cite{Juan,Juan2,Cristiano,MH,Vicente,PHM,Ballesteros,Ballesteros2}. While the amplitude of perturbations on cosmological scales is of order $P_\Phi\sim 10^{-9}$, the r.m.s. amplitude at the bump should be much larger, $P_\Phi\sim 10^{-3}-10^{-2}$, so that PBHs can form in significant abundance.
This is a strong departure from scale invariance, and it seems plausible that in generic models of this sort the power spectrum might be enhanced also at the scales interpolating from the cosmological plateau down to the PBH scale, including the intermediate scales comparable to the co-moving size of the binaries. Here, we shall be agnostic about the specific inflationary dynamics, and will simply explore the consequences of an enhanced spectrum which we shall model as a (piecewise) power law $P_\Phi(k) \propto k^p$. As we shall see, such an enhancement may have potentially observable consequences. In particular, it may affect the differential merger rate of binaries as a function of the component masses.

The paper is organized as follows. In Section \ref{general} we briefly review the formation of PBH binaries and the distribution of orbital parameters, taking into consideration the effect of neighbouring black holes but ignoring cosmological perturbations. It is in this context that the universality coefficient $\alpha$, which charaterizes the dependence of the merger rate on the masses of the components, was first introduced \cite{Kocsis}. Hence, this will be a useful reference case. 
We also comment on PBH infalls and their effect on $\alpha$. 

In Section \ref{cosmological} we discuss the effect of cosmological perturbations on the dimensionless orbital angular momentum parameter $j$. In contrast with earlier analysis, here we include the perturbations in radiation, whose effect dominates over that of matter perturbations for binaries which decouple from the Hubble flow deep in the radiation era. For an enhanced $P_\Phi$ these tend to dominate the distribution of $j$.

In Section \ref{mergerrates} we consider the merger rates in three different scenarios: the nearly scale invariant cosmological plateau (Case A), an enhanced spectrum at intermediate scales with a moderate spectral index $0<p\lesssim 2$ (Case B), and a rather steep power spectrum $p>2$, peaked at scales smaller than the binary size (Case C). Our conclusions are summarized in Section \ref{conclusions}. 

Throughout the paper $f$ will denote the fraction of dark matter in the form of PBHs, and $s$ will denote the cosmological scale factor, while $a$ will denote the semi-major axis of binaries. We adopt the convention that $s=1$ at the time of matter-radiation equality. The speed of light is set to $c=1$.

\section{PBH binary formation and universality}
\label{general}

In this Section, we briefly review the case where  the angular momentum of binaries is due to the tidal torque 
from other PBHs in the vicinity, neglecting cosmological perturbations. We also introduce the universality coefficient $\alpha$ \cite{Kocsis}, and we discuss how this may be affected by the infall of neighboring PBH on binaries.

\subsection{Initial orbital parameters and the life-time of binaries}

Following \cite{Sasaki1,Thorne}, let us assume a uniform distribution of PBHs, without any initial spatial correlations.\footnote{The effect of such an initial correlation has been discussed in Refs. \cite{Guillermo,Domcke}.}
From a given PBH, the probability of finding the nearest neighbour at a certain distance is given by
\begin{equation}
dP = e^{-X} dX. \label{xdist}
\end{equation}
Here $X = n V$ is the product of the co-moving number density $n$ times the co-moving volume $V=(4/3)\pi x^3$, where $x$ is the co-moving distance. We adopt the convention that the cosmological scale factor is 
$s=1$ at the time of matter-radiation equality. 

We shall also assume that the PBH mass function is not too broad \footnote{This is expected when PBH are formed from very high peaks of a Gaussian random field of density perturbations \cite{GM,YHGK}, even if the enhancement in the power spectrum has a sizable width. Unless the power spectrum involves different explicit scales, high peaks of the random field tend to have a well defined shape, which leads to a relatively narrow range of masses after gravitational collapse.}, allowing however for some spread in the masses within an order of magnitude or so. 
The co-moving number density takes the form $n=f \rho_{eq}/(2\bar m)$, where $f$ is the fraction of DM in the form of PBHs, $\rho_{eq}$ is the density at the time of equality, and $\bar m$ is the average mass in the distribution. 
We may then write
$$
X = n V = \left({x\over \bar x}\right)^3,
$$
where
\begin{align}
    \bar{x}=\left(\frac{3\bar m}{2\pi f \rho_{eq}}\right)^{1/3}. \label{barx}
\end{align}
In a spherical region of radius $\bar x$ we expect to find one PBH, on average, so the length scale $\bar x$ can also be thought of as a typical separation between PBHs.

A pair of black holes forms a binary when the relative kinetic energy due to the Hubble flow becomes comparable to the gravitational binding energy between the two objects  \cite{Sasaki1},
\begin{equation}
{1\over 2} \mu H^2 s^2 x^2 \sim {G m_1 m_2 \over  s x}. \label{formation}
\end{equation}
Here, $\mu = m_1 m_2/M$ is the reduced mass, where $M =m_1+m_2$ is the total mass of the binary. The above relation  has to be satisfied before the end of the radiation era, 
since both sides will scale as $s^{-1}$ during matter domination. For $x\lesssim \bar x$, and taking into account that $\rho \approx \rho_{eq}/(2s^4)$ in the radiation era ($s\ll1$), the relation (\ref{formation}) 
is satisfied when the cosmological scale factor $s$ is of order
$ s \sim \lambda (\bar m/M)\leq1,$
where we have introduced
\begin{equation}
\lambda\equiv {X\over f}.\label{lambdadef1}
\end{equation}
More precisely,  parametrizing the physical distance as $d = \chi(\eta;\lambda) x_0$, where $x_0$ is the initial co-moving separation, we may write
\begin{equation}
x=|\vec x| = {\chi(s)\over s} x_0. \label{timedependence}
\end{equation}
The numerical analysis in Refs. \cite{Kam,RSVV} shows that, that for binaries forming deep in the radiation era (i.e. $\lambda\ll 1$), the function $\chi$ is self-similar
\begin{equation}
\chi(s;\lambda) = \lambda \chi(s/\lambda;1).
\end{equation}
Initially, the two PBH are following the Hubble flow, so that $\chi \approx s$, $\vec x \approx \vec x_0$ is approximately constant, and the physical distance grows linearly in $s$. However, when the scale factor reaches the value 
\begin{equation}
s\approx s_b 
={\lambda\over 3}\left({2\bar m\over M}\right), \label{sb}
\end{equation} 
the physical distance turns around and a bound system is formed with semi-major axis given by \cite{Kam}
\begin{equation}
a 
= 3\beta s_b {x\over 2} = {\beta\over 2} \left({2\bar m\over M}\right)\Big(\frac{3\bar m}{2\pi \rho_{eq}}\Big)^{1/3}\lambda^{4/3}, \label{a}
\end{equation}
where $\beta\approx 0.2$. 
Introducing the dimensionless average mass parameter 
\begin{equation}
m\equiv {\bar m\over M_\odot}.
\end{equation}
we have
\begin{equation}
a \approx 1.8\cdot 10^{-7} \lambda^{4/3} m^{1/3} \left({2\bar m\over M}\right) H_{eq}^{-1}, \label{semimajor}
\end{equation}
where we have used $H_{eq}^{-1} \approx 0.9\cdot 10^{18} km$. 

       
In an environment with no external forces and torques, two PBHs which are initially at rest would collide head-on due to gravitational attraction in a very short time-scale 
\begin{equation}
\Delta t \sim a^2(GM a)^{-1/2} \sim H_b^{-1} \lesssim t_{eq}, \label{deltat} 
\end{equation}
comparable to the Hubble radius $H_b^{-1}$ at the time when the binary forms. 
However, the binary system is immersed in a local tidal field, created by density perturbations and by other PBHs in the neighborhood.  These forces will exert a torque on the binary, giving it an orbital angular momentum which avoids the head-on collision. The binary will then slowly radiate its energy by emitting gravitational waves in a much longer timescale, before the final merger occurs. 

For a binary with initial orbital angular momentum $\ell$ per unit reduced mass, the life-time is given by Peters formula
\cite{Peters} 
\begin{equation}
t=t[j,a]\equiv {3\over 85}{a^4\over G^3 m_1 m_2 M} j^7, \label{petersf}
\end{equation}
where dimensionless parameter $j$, is defined as
\begin{equation}
j \equiv {\ell \over \sqrt{GMa}}.\label{jandell}
\end{equation}
For an elliptic orbit with semi-minor axis $b$, we have $j = b/a = \sqrt{1-e^2}$, where $e$ is the eccentricity. Note that $j<1$. 
Using $t=t_0 \approx 1.3 \cdot{10^{23}}~km$, we find that the binaries which are merging today are characterized by
\begin{equation}
j=j_0(\lambda)\approx 1.6 \cdot{10^{-3}} \lambda^{-16/21} m^{5/21} \left({M\over 2\bar m}\right)\left({4m_1m_2\over M^2}\right)^{1/7}. \label{j0}
\end{equation}
where we have used Eq. (\ref{a}) with $\beta\approx 0.2$.
The distribution of $j$ in the ensemble of binaries depends on the specific mechanisms which give the binaries their angular momentum.

\subsection{Merger rates}

In general, the differential number density of binaries per unit volume is given by
\begin{equation}
dn_{bin}= dn_M d^2{\cal F}.
\end{equation}
Here 
\begin{equation}
d^2{\cal F}\equiv d{\cal F}(m_1) d{\cal F}(m_2),
\end{equation}
where $d{\cal F}$ is the PBH mass distribution function, and $dn_M(m_1,m_2,X,j)$ is the distribution of binaries with masses $m_1$ and $m_2$, initial separation of the partners characterized by $X$ and orbital angular momentum parameter $j$.  The variable $X$ is distributed as (\ref{xdist}), so using $X=f\lambda$, we have
\begin{equation}
dn_M= \Theta\left(M- \lambda \bar m\right)\ f^2 {\rho_m(t_0)\over 2\bar m} e^{-f \lambda} dP(j; \lambda) d\lambda,
\end{equation}
where $dP(j;\lambda)$ is the distribution of $j$ for given $\lambda$, $f \rho_m(t_0)/\bar m$ is the number density of PBH at the present time, $\rho_m$ is the current matter density, and we have inserted a factor of $1/2$ to avoid double counting of binaries.
The Heavyside function restrics the range of $\lambda$ since, according to our earlier discussion around Eq. (\ref{formation}), a given PBH will only be part of a binary if the distance to the nearest PBH satisfies
\begin{equation}
\lambda\lesssim {M\over \bar m}.
\end{equation}
Otherwise the Hubble flow velocity always remains larger than the binding energy. If the distribution of masses is not too wide, we have $M \sim 2\bar m$, and 
for $f \ll 1$ the exponential factor $e^{-f \lambda}$ can be approximated by $1$.

The intrinsic merger rate\footnote{Here, and for the rest of this paper, we consider intrinsic merger rates, ignoring effects due to time delay of events which occur at high redshift. 
These can be incorporated along the lines of  Ref.  \cite{RSVV}.}  of PBH binaries per unit time and volume can be written as
\begin{equation}
d\Gamma(t_0,m_1,m_2) = \Gamma_M(t_0) d^2{\cal F} \label{rac1}
\end{equation}
where the rate at fixed total mass $M=m_1+m_2$ is given by
\begin{equation}
\Gamma_M(t_0) = \int \delta\left(t_0-t[j,a]\right) dn_M.\label{ratefor}
\end{equation}
Using (\ref{petersf}) in the argument of the delta function, we can perform the $j$ integration to obtain
\begin{equation}
\Gamma_M(t_0) = f^2 {\rho_{m}\over 14\bar m t_0} \int_{\lambda_{min}}^{M/\bar m}  W(\lambda) d\lambda. 
\label{intrate}
\end{equation}
Here ,we have dropped the factor $e^{-f \lambda}$, since as mentioned above this can be approximated by unity in the relevant range of parameters.
Also, we have introduced
\begin{equation}
W(\lambda) =  j \left.{dP(j;\lambda) \over dj}\right|_{j=j_0}, \label{W}
\end{equation}
where $j_0$ given by Eq. (\ref{j0}) is the value of $j$ for which the lifetime of the binary coincides with the present age of the universe $t_0$.
The integral in (\ref{intrate}) is in the range
\begin{equation}
\lambda_{min} < \lambda \lesssim M/\bar m, \label{lambdarange}
\end{equation}
where the lower limit\footnote{In the expressions which are given only by order of magnitude, and in the interest of brevity, we will often omit the explicit dependence on the individual masses, assuming they are within one order of magnitude or so from each other.} $\lambda_{min}\sim 2 \cdot 10^{-4} m^{5/16}$ is determined from (\ref{j0}), taking into account that we must have $j_0 \leq 1$.

The distribution $dP(j;\lambda)$ depends on the mechanism which gives angular momentum to the binaries. In general, 
\begin{equation}
j= |\vec \jmath_{nb} + \vec \jmath_{cp}|
\end{equation}
is the added contribution from torques due to neighboring PBHs, and from torques due to cosmological perturbations. For the rest of this Section we concentrate on $j_{nb}$, while the effect of cosmological perturbations will be discussed in the following Sections.

\subsection{Tidal torque due to neighboring black holes} \label{ssthird}

Let us start by considering the effect of a single neighboring PBH, producing a tidal torque on the binary. 
By integrating the torque over time, we have,
\begin{equation}
\vec \jmath\sim (\vec x \times \Delta \vec g) {\Delta t \over \sqrt{GMa}}, \label{jest}
\end{equation}
where $\Delta t$ is given in Eq. (\ref{deltat}). Here, and for the rest of this section, we suppress the subscript $nb$ from $\vec \jmath$, since we are only dealing with the effect of neighboring PBHs. 
In Eq. (\ref{jest}),  $\vec x=\vec x_2-\vec x_1$ is the relative co-moving separation between the members of the binary and 
\begin{equation}
\Delta \vec g = \vec\nabla \Phi(\vec x_1) - \vec \nabla\Phi(\vec x_2),
\end{equation}
is the tidal acceleration, expressed in terms of the Newtonian potential $\Phi$ created by the neighboring PBH. If the gradients vary on a length scale much larger than the separation $\vec x$, the tidal acceleration can be expanded 
in powers of $\vec x$ and the leading term is given by $\Delta g^k = - \Phi,_{kl} x^l$,
where the spatial derivatives are with respect to the co-moving coordinates.  If the 3rd black hole has a mass $m_3$, and is at a co-moving distance $y\gtrsim x$, then 
$\Phi\approx Gm_3/(s_b x)$, where $s_b$ is the scale factor around the time when the binary forms. The co-moving gradients can then be estimated as 
$|\Phi_{,ij}| \sim (Gm_3/s_b y^3) (3 y^i y^j -\delta_{ij})$, and Substituting in (\ref{jest}) with $a\sim s_b x$ we obtain
\begin{equation}
j= |\vec\jmath|=
\gamma {j_X\over Y}, \label{jsim}
\end{equation}
Here we have introduced the variable $Y\equiv (y/\bar x)^3$, characterizing the distance to the nearest third PBH, and 
\begin{equation}
j_X=(\bar m/M) X. \label{jx}
\end{equation}
Since the position of the third black hole is random, the distribution for $Y$ is also given by
\begin{equation}
dP(Y)= e^{-Y} dY. \label{ydistri}
\end{equation}
The coefficient $\gamma$ is given by 
\begin{equation}
\gamma \approx 1.5\ |\sin2\theta_3|{m_3\over \bar m},
\end{equation}
where $\theta_3$ is the angle between the relative coordinate $\vec x$ and the position of the third black hole $\vec y$, that is, $\cos \theta_3 = \hat x\cdot \hat y$.
The overall numerical factor in this expression is determined by taking into consideration the time dependence in Eq. (\ref{timedependence}) as the binary forms \cite{Kam,Ioka,Kocsis,RSVV}.
Note that the average of $\gamma$ over angle and mass distribution function is 
$$
\bar\gamma \approx 1. 
$$
Following \cite{Kocsis}, we keep $\gamma$ as an undetermined random variable of order one, over which we can integrate at the end of the computation, if needed.

\begin{figure}
\begin{center}
\includegraphics[scale=0.7]{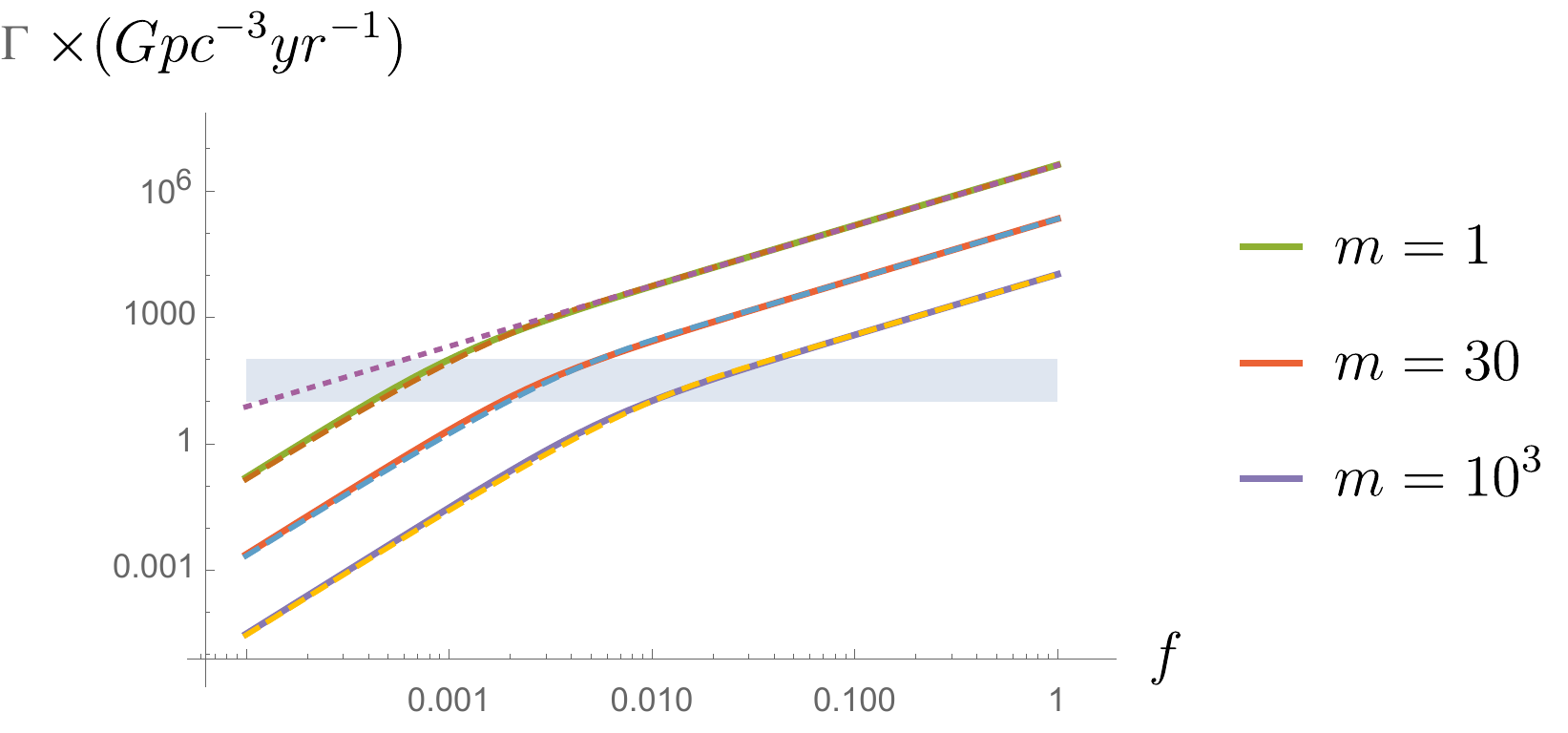}
\caption{The dashed lines represent the merger rate $\Gamma^{(1)}_M$ of PBH binaries [given by Eq. (\ref{intrate}) with Eq. (\ref{justone})], as a function of $f$ for different values of the mass. This estimate assumes that the initial angular momentum is due to the closest neighbouring black hole, and
naively counts all binaries with the appropriate initial conditions for merging at the present time, as if they were in complete isolation. The thick lines represent the merger rate 
$\Gamma^{(\infty)}_M$ [given by Eq. (\ref{intrate}) with Eq. (\ref{many})], where torques from all neighboring PBH are included. Solid and dashed curves nearly coincide, in agreement with the notion that it is the closest PBH that gives the dominant contribution to the torque.
The gray shaded region corresponds to the merger rates observed by LIGO/Virgo.
The approximation Eq. (\ref{rateflarge}) is also shown as a dotted line for $m=1$. Unless otherwise stated, we will use $m_1=m_2=\bar m$ in all figures.
\label{undisrup}}
\end{center}
\end{figure}

Taking into consideration (\ref{jsim}) and (\ref{ydistri}), we have
\begin{equation}
dP^{(1)}(j;\lambda) = \gamma{j_X\over j^2} \exp\left({-\gamma{j_X\over j}}\right) dj. \label{jdistrione}
\end{equation}
Here, the superindex in the probability distribution indicates that, for the time being, we are considering the effect of the nearest neighboring PBH only.
Then, we have
\begin{equation}
W^{(1)}(\lambda)= j \left.{dP^{(1)}(j;\lambda)\over dj}\right|_{j=j_0} = Y_0(\lambda) e^{-Y_0(\lambda)}, \label{justone}
\end{equation} 
where
\begin{equation}
Y_0(\lambda) \equiv\gamma {j_X\over j_0(\lambda)}=\gamma \left({\lambda\over \lambda_*}\right)^{37/21}.
 \label{Y01}
\end{equation}
The characterisctic value 
\begin{equation}
\lambda_* \approx 3.7 \cdot 10^{-2} f^{-{21/37}} m^{5/37} \left({M\over 2\bar m}\right)^{42/37} \left({4m_1m_2\over M^2}\right)^{3/37}.
\end{equation}
is essentially the peak of the function $W^{(1)}(\lambda)$. The rate (\ref{intrate}) is plotted in Fig. \ref{undisrup} for a range of values of $f$, and for different values of the mass $\bar m$, assuming that $m_1=m_2=\bar m$ and $\gamma\approx 1$. The curves have a knee which separates two different regimes with a power law behaviour in $f$. This can easily be understood analytically. The behaviour of the merger rate depends on whether $\lambda_*$ is 
large or small, and this in turn depends on the value of $f$. 
 
The behaviour of the merger rate below the knee ($\lambda_* \gg1$) corresponds to a low fraction of DM in PBH, $f\ll f_*$, where
\begin{equation}
f_*\sim 3\cdot 10^{-3} m^{5/21}. 
\label{smallf}
\end{equation}
Since $\lambda\lesssim M/ \bar m \sim 1 \ll \lambda_*$, we have $Y_0\ll 1$, 
and we may then neglect the exponential dependence of the integrand in (\ref{intrate}),
\begin{equation}
\int_0^{M\over \bar m}  d{\lambda} W^{(1)} \approx  \gamma \int_0^{M\over \bar m} \left({\lambda\over \lambda_*}\right)^{37/21}d\lambda .
\end{equation}
Since the lower limit of integration $\lambda_{min}\sim 10^{-4}$ does not play a role, we have set it to zero for simplicity. 
This leads to the estimate
\begin{equation}
\Gamma_M^{(1)} \sim {1.6 \cdot 10^{11}\over Gpc^3 yr} \gamma f^3 m^{-26/21}\left({M\over 2\bar m}\right)^{16/21} \left({M^2\over 4 m_1m_2}\right)^{1/7}, \quad ({\rm low}\ f)\label{rateflow}
\end{equation}
where we have used
\begin{equation}
{\rho_{m}\over M_\odot t_0} \approx 3\cdot 10^9 Gpc^{-3} yr^{-1}. \label{numerics}
\end{equation}
For an approximately monochromatic PBH mass function, the observational bound $\Gamma \lesssim 10^2  Gpc^{-3} yr^{-1}$ then leads to 
\begin{equation} 
f\lesssim 0.85 \cdot 10^{-3} m^{26/63}. \label{boundonf}
\end{equation}
We conclude that, if the third black hole is the dominant source of orbital angular momentum, then solar mass black holes, with $m\sim 1$, can only account for a very small fraction of dark matter, with $f\lesssim 10^{-3}$. This is in agreement with the analysis of  Refs. \cite{Sasaki1,Kocsis}. Note that, even in the case $m\sim 1$ the upper limit of the observational bound (\ref{boundonf}) satisfies the condition $f\lesssim f_*$ only marginally. 


Hence, let us now consider the complementary limit $f\gtrsim f_*$.
For $\lambda \gg \lambda_*$ we have  $Y_0(\lambda)= \gamma(\lambda/\lambda_*)^{37/21} \gg1$, and due to the factor $e^{-Y_0}$ the integral (\ref{intrate}) is effectively cut-off at $\lambda=\lambda_*\ll1$. Therefore it is a good approximation to remove the upper limit of integration, which doesn't play a role, and then the integral scales as $\lambda_*$,
\begin{equation}
\int_0^{M\over \bar m}W^{(1)} d\lambda \approx \int_0^{\infty} d\lambda Y_0 e^{-Y_0} \approx {21\over 37} \Gamma\left({58\over 37}\right) \gamma^{-21/37}\lambda_*.\label{lambdascaling0}
\end{equation}
The rate can then be approximated as
\begin{equation}
\Gamma_M^{(1)} \sim {4 \cdot 10^6 \over Gpc^{3} yr}f^2 (\gamma f)^{-21/37} m^{-32/37} \left({M\over 2\bar m}\right)^{42/37}\left({4m_1m_2\over M^2}\right)^{3/37}. \quad ({\rm higher}\ f)\label{rateflarge}
\end{equation}
The observational bound $\Gamma \lesssim10^2  Gpc^{-3} yr^{-1}$ then leads to the condition
\begin{equation}
f\lesssim 0.6 \cdot 10^{-3} m^{32/53}, \label{boundonf2}
\end{equation}
where, as in Eq. (\ref{boundonf}), in this inequality we assume a nearly monochromatic PBH mass function.
In the mass range of LIGO/Virgo detections, $m \sim 30$, the bound on the fraction $f$ of DM in PBH is limited to $\sim 0.5 \%$, again in good agreement with \cite{Sasaki1,Kocsis}. The analytic estimate (\ref{rateflarge}) is plotted in Fig. \ref{undisrup} as a dotted line for $m=1$.

The previous considerations can be extended to the case where we include the torque of all neighboring black holes, and not just the closest one.
An expression for $dP^{(\infty)}(j)$ due to the cumulative effect of all PBHs in the neighborhood was derived in \cite{Kam,RSVV,Kocsis}. This has the form of a power law distribution with a break at  $j_X \equiv (\bar m/M) X = (\bar m/M) f\lambda$:
\begin{equation}
j{dP^{(\infty)}(j;\lambda)\over dj} = {(j/j_X)^2\over (1+(j/j_X)^2)^{3/2}}. \label{break}
\end{equation}
Note that at large $j$, the behaviour of (\ref{break}) is similar to (\ref{jdistrione}), where only the nearest PBH is considered. However, at small $j$ the distribution (\ref{jdistrione}) vanishes exponentially in $1/j$, while (\ref{break}) has 
the form $dP^{(\infty)} \propto j dj$. As pointed out in \cite{Kocsis}, the reason is that in the case of a single PBH, the only way to reduce the torque on the binary is to place the PBH sufficiently far. The probability 
for that decays exponentially in $1/j$ for large distance. On the other hand, when many neighboring PBHs are involved, their added torques may randomly produce a small effect, with a probability which is only phase space suppressed. The dimensionless angular momentum $\vec \jmath$ is in the plane orthogonal to the initial relative separation $\vec x$, so the corresponding measure is two dimensional 
$d^2\vec \jmath = 2 j dj$, and the behaviour $dP\propto jdj$ is expected. In conclusion, the distribution (\ref{jdistrione}) does not provide a very good description at small $j$, even if it is true that the nearest PBH gives the dominant contribution to the torque.

Using (\ref{break}) in (\ref{W}) we have
\begin{equation}
W(\lambda) = {\bar Y_0 \over (1+ \bar Y_0^2)^{3/2}}, \label{many}
\end{equation}
where 
\begin{equation}
\bar Y_0 = {j_X\over j_0} = \left({\lambda\over\lambda_*}\right)^{37/21}, \label{bary0}
\end{equation}
is the same as $Y_0$ given in (\ref{Y01}), with the coefficient $\gamma$ replaced by its averaged value over masses and directions, $\bar \gamma=1$.

At low $f$, where $\lambda_*\gg 1$, we have $\bar Y_0 \ll 1$ throughout the range of integration in (\ref{intrate}). Hence, $W\approx \bar Y_0$, and the rate will be given by Eq. (\ref{rateflow}). In other words, the inclusion of 
the effect of an infinite number of neighbours does not change the merger rate: 
\begin{equation}
\Gamma_M^{(\infty)} \approx \Gamma_M^{(1)},\quad ({\rm low}\ f)
\end{equation}
Also, at high $f$, where $\lambda_* \ll 1$, the integral is dominated by $\lambda \sim \lambda_* \ll M/\bar m$, so we can approximate by extending the range of integration to infinity and evaluating in terms of Euler's Gamma function. 
Then one finds
\begin{equation}
\Gamma_M^{(\infty)} \approx 0.95 \Gamma_M^{(1)},\quad ({\rm higher}\ f)
\end{equation}
and the difference between the two cases is only by a very small change in the overall numerical factor. In fact, the distributions (\ref{justone}) and (\ref{many}) produce integrated merger rates which are almost indistiguishable from one another (also for intermediate values of $f\sim 10^{-3}$), in agreement with the notion that the nearest PBH gives the dominant contribution to the torque (See Fig. \ref{undisrup}). More importantly, the dependence of the merger rate on binary masses is basically unaffected by the inclusion of an infinite number of neighbours. Let us now turn to the characterization of such mass dependence.

\subsection{Universality in the mass dependence of the merger rates.}

In principle, we cannot predict the mass dependence of the merger rates unless the initial mass distribution function $d{\cal F}(m_i)$ is known. Unfortunately, the latter is model dependent. However, a very interesting observation was made in Ref. \cite{Kocsis} which may bypass this difficulty.
Noting that the rate in a given mass interval $\Delta m_1$, $\Delta m_2$ is given by $\Delta\Gamma(m_1,m_2) =\Gamma_M(m_1,m_2)\Delta{\cal F}(m_1)\Delta{\cal F}(m_2)$, the expression
\begin{equation}
\alpha \equiv -M^2 {\partial^2 \over \partial m_1 \partial m_2} \ln[\Delta\Gamma(m_1,m_2)] \label{alphadef}
\end{equation}
is independent of the unknown distribution funcion ${\cal F}$. 
It was argued in \cite{Kocsis} that with a sufficiently large sample of PBH merger events, of order $10^3$, the coefficient $\alpha$ can be determined observationally with accuracy of order $15\%$. 
This makes it a very attractive observable, within reach of existing and upcoming gravitational wave detectors \cite{review}.

If neighboring PBHs are the only source of angular momentum for the binaries, the coefficient $\alpha$ can readily be found from the expressions (\ref{rateflow}) and (\ref{rateflarge}).
The powers of $m_1m_2$ in these expressions for the merger rate do not contribute to $\alpha$, since after taking the logarithm and the two derivatives with respect to $m_1$ and $m_2$ such terms drop out. This is the same reason why $\alpha$ does not depend on the initial mass distribution functions $\cal F$. The only contributions to $\alpha$ come from powers of the total mass $M$.
Hence, from (\ref{rateflow}), we have 
\begin{equation}
\alpha= 22/21,\quad ({\rm low} f)
\end{equation}
 and from (\ref{rateflarge}) we have 
\begin{equation}
\alpha=36/37. \quad({\rm higher} f) \label{alphalarge}
\end{equation}
This leads to the prediction of a ``hidden universality" in the merger rate \cite{Kocsis}, where the parameter $\alpha$ should be in the narrow range 
\begin{equation}
0.97 \lesssim \alpha \lesssim 1.05. \label{unitorque} 
\end{equation}
As we shall see, this universality coefficient can be altered by different effects [see Fig. \ref{alpha}]. This may convey useful information about the actual circumstances surrounding binary formation
and evolution.

\subsection{PBH infalls and the universality coefficient}

\begin{figure}
\begin{center}
\includegraphics[scale=0.7]{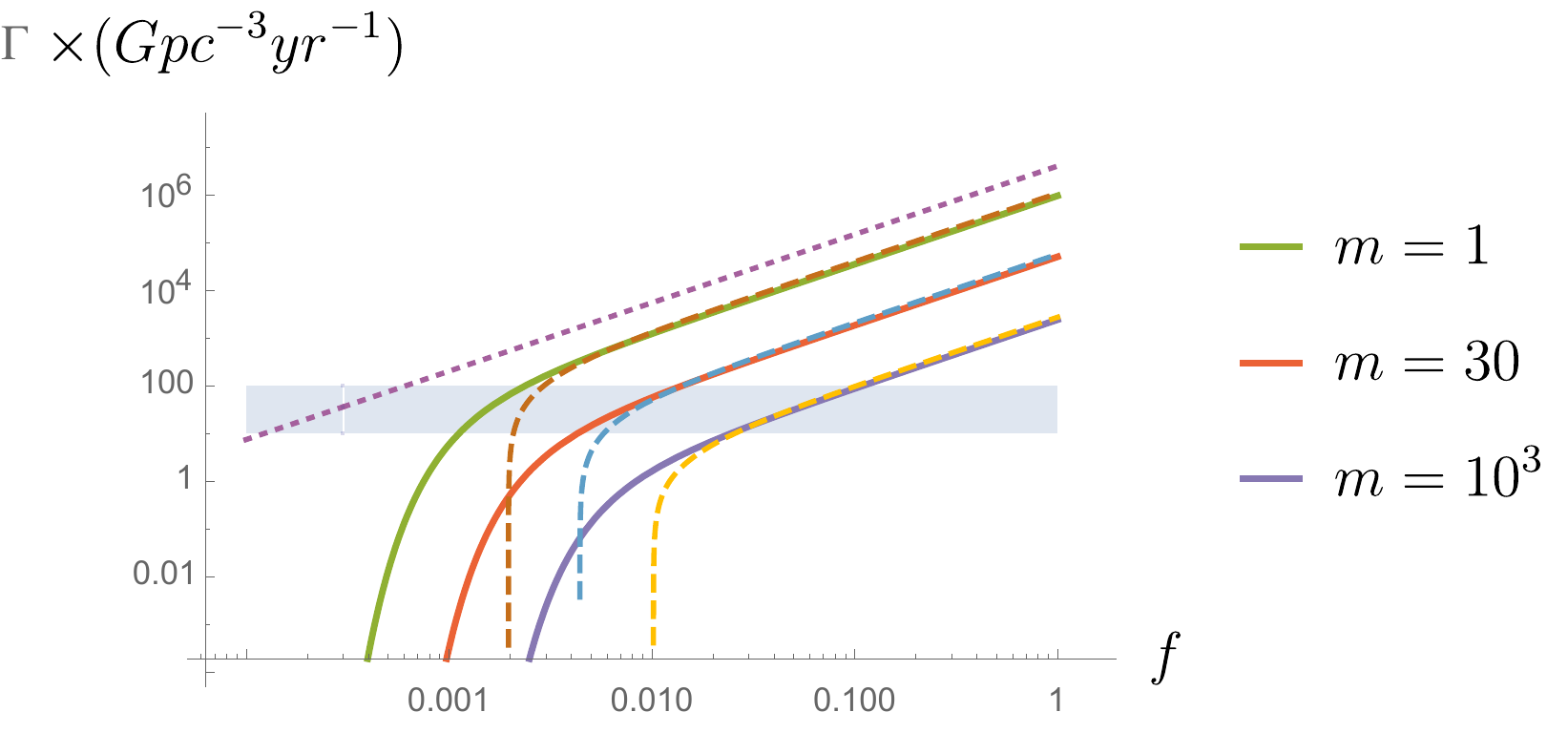}
\caption{The effect of PBH infalls onto binaries. The plot is the same as Fig. \ref{undisrup}, but now excluding all binaries with $Y< Y_{min}=2$ since these are likely to be disrupted by the infall of neighbouring PBHs. For comparison, the approximation Eq. (\ref{rateflarge}) is also shown as a dotted line for $m=1$. Dashed lines correspond to the crude approximation where only the closest PBH contributes to the torque, whereas solid lines include the torque from all neighboring PBHs. Generally, the rate is suppressed at large $f$ by a factor of order 
$e^{-{Y_{min}}}$. 
At low $f$ the rate drops rather dramatically since it is hard to give the binaries the necessary angular momentum unless there is a neighbour close enough to the binary. However, as we shall see, this dramatic drop is avoided when the torque from cosmological perturbations is included (see e.g. Fig. \ref{caseA}).}
\label{undisrup2}
\end{center}
\end{figure}

Eq. (\ref{intrate}) includes all binaries which have the appropriate initial conditions to merge at the present time, provided that they remain in isolation for the rest of cosmic history. 
It has been argued in \cite{RSVV} that this overestimates the rate, because some of the binaries may be affected by the infall of neighboring PBHs, which could disturb the eccentricity
or even disrupt the binary. Note that binaries merging at the present time have a very low $j$, which is given by Eq. (\ref{j0}), $j\sim 10^{-3}$. PBH infalls would increase this value by a large factor,
making the life-time of the binary much larger than the age of the universe.
On the other hand, if the closest neighbour is at a distance such that $Y\gtrsim M/\bar m$, then the binary is in an underdense region, and the 
neighbouring PBHs will not decouple from the Hubble flow to fall onto it \cite{Ioka,RSVV}.

To illustrate the potential impact of infalls as a function of $f$, let us start by using a crude approximation where only the closest PBH contributes to the torque. Within this approximation, we may compare the naive rate given by Eq. (\ref{intrate}) with the merger rate of ``pristine" binaries which are unaffected by infalls. This is achieved by restricting the integration to the range where, say, 
\begin{equation}
Y> Y_{min} = {M\over \bar m}. \label{ymin}
\end{equation} 
The result is illustrated in Fig. \ref{undisrup2} (dashed lines). For that comparison, we assume a monochromatic mass spectrum, and the lower limit of the integral in Eq. (\ref{intrate}) is taken to be the value of $\lambda$ that corresponds to $Y_0(\lambda)=M/\bar m =2$. 
For very low $f\sim 10^{-3}$, the lower limit of integration becomes of order one, and for lower $f$ the range of integration completely disappears. This causes the sudden drop of the dashed curves in Fig. \ref{undisrup2}. 
On the other hand the effect is not so dramatic for higher $f\gtrsim 10^{-3}$, amounting only to a moderate overall factor of order 10 or so. The reason is simple to understand. According to the analysis of Subsection \ref{ssthird}, for $f$ above the knee in the curves of Fig. \ref{undisrup}, the rates are dominated by binaries where the closest neighbor is at a
distance $Y_0(\lambda_*) \sim 1$. In this case the factor $e^{-Y_0(\lambda)}$ acts as the effective cut-off of the integral (\ref{intrate}) at the value $\lambda_*\lesssim 1$. Hence, the effect of restricting the range of integration to $Y>Y_{min}=2$ is only a mild suppression by a factor of order $\exp[-(Y_{min}-Y(\lambda_*))]$, where the exponent is of order 1.

Since PBH infalls have a sizeable impact on the merger rates, it important to consider the overall effect of all neighboring PBH under the assumption that all of them are outside the basin of attraction of 
the binary. This issue was considered in generality in Ref. \cite{RSVV}, where the distribution $dP(j;\lambda,Y_{min})$ was calculated, by taking into account {\em all} neighboring PBHs, and assuming that these are at a distance larger than $Y_{min}$ from the binary. The result was found in closed form in terms of integrals of Hypergeometric functions, and it is somewhat cumbersome in general. However, the expression greatly simplifies in the limits $Y_{min}\to 0$ and $Y_{min}\gg  1$, which are of our primary interest. For $Y_{min} \to 0$, Eq. (\ref{break}) is recovered, as expected, since in this case we do not exclude any of the binaries from the count.  On the other hand, for $Y_{min}\gg 1$, it was found that the distribution can be approximated as a Gaussian 
\begin{equation}
dP(j;\lambda,Y>Y_{min}) \approx  \exp\left(-{j^{2}\over \sigma_{nb}^2}\right) {2j dj\over \sigma_{nb}^2}, \quad (Y_{min} \gg 1) \label{gauss} 
\end{equation}
with variance
\begin{equation}
\sigma^2_{nb} ={K \over Y_{min}}  j_X^2. \label{sigmanb}
\end{equation}
Here, $K=6\langle m^2\rangle/ (5\bar m^2)\sim 1$, where the brackets indicate average over the mass distribution.
The approximation is already quite accurate for $Y_{min} \gtrsim 2$, which is the range of our interest.

The differential merger rate is therefore given by
\begin{equation}
W(\lambda)=\left.j{dP(j;\lambda,Y>Y_{min}) \over dj}\right|_{j=j_0} e^{-Y_{min}}
,
\end{equation}
where the factor $e^{-Y_{min}}$ accounts for the probability that the closest neighboring PBH is further than $Y_{min}$.
Here, and in what follows, we restrict attention to the regime where $f \gtrsim f_*$, which seems most relevant for observations. 
Then, we can approximate (\ref{intrate}) by removing the upper limit of integration and we have
\begin{equation}
\int_0^{\infty} W d\lambda \approx {21\over 37}\Gamma\left({53\over 74}\right)(K\bar\gamma^2)^{-21/74}  \lambda_* Y_{min}^{21/74} e^{-Y_{min}}.\label{complete}
\end{equation}
where we have used (\ref{bary0}).  The corresponding merger rate is plotted in Fig. \ref{undisrup2} (solid lines), for different values of the mass.

Let us now consider the universality coefficient $\alpha$. If $Y_{min}$ were independent of the masses, then (\ref{scalingsingle}) would scale like $\lambda_*$, just like in Eq. (\ref{lambdascaling0}). In that case, we would recover the value $\alpha = 36/37$. However, since heavier binaries have a larger basin of attraction, $Y_{min}$ scales as (\ref{ymin}),  and we have\footnote{An analogous computation considering only the effect of the nearest PBH produces a somewhat different answer,
\begin{equation}
\int W d\lambda \approx \int Y_0  e^{-Y_0} d\lambda \approx {21\over 37} \gamma^{-{21\over37}} \lambda_* \int_{Y_{min}}^{\infty} Y_0^{21\over37} e^{-Y_0}dY_0 = {21\over 37} \gamma^{-{21\over37}} \lambda_*
\Gamma(58/37,Y_{min}).\label{scalingsingle}
\end{equation}
Approximating
$\Gamma(58/37,Y_{min})\approx 1.25\ Y_{min}^{21/37} e^{-Y_{min}}$  for $Y_{min} \gtrsim 2$,
and using (\ref{ymin}) we find $\alpha = (36+21)/37\approx 1.54$. However, it should be noted that here we are considering large $Y$, which corresponds to small $j$, and in this regime it is not a good approximation to neglect the 
contribution from all other PBHs, as explained in the paragraph following Eq. (\ref{break}).}
\begin{equation}
\alpha= {36 \over 37}+{21\over 74} \approx 1.26. \label{alphamany}
\end{equation}
The second term comes from $M^{21/74}$ in (\ref{complete}). It is interesting to note that the factor $e^{-Y_{min}}$, which is exponential in $M$, does not contribute to the parameter $\alpha$, since it drops out after taking two derivatives of $\log \Gamma$ 
with respect to the masses. We conclude that the effect of infalls of neighboring PBHs onto binaries produces a significant shift 
of $\alpha$ towards a higher value.
\footnote{Aside from infalls, the simulations in Ref \cite{RSVV} also indicate that, for $f\gtrsim 0.1$, binaries can be disrupted during the matter dominated era by interaction with compact N-body systems. This effect can be particularly important for $f\sim 1$, where a sizable fraction of the binaries undergo interactions even before the time of recombination. This effect is also likely to suppress the rates at high $f$, and further work is needed to assess what fraction of the binaries may ultimately remain unaffected. In what follows, we shall simply ignore this possibility, assuming that $f$ is low enough for this effect to be unimportant.}
Let us now turn our attention to the effect of cosmological perturbations. As we shall see, the parameter $\alpha$ can also be sensitive to these.

\section{Cosmological perturbations}
\label{cosmological}

In this Section we consider the effect of cosmological perturbations on the eccentricity of PBH binaries. In contrast with previous work, we include the effect of density perturbations in radiation, which can be dominant for binaries with $\lambda\ll1$, decoupling from the Hubble flow deep in the radiation era.

Let us assume that primordial density perturbations are adiabatic and Gaussian. These are completely characterized by the primordial power spectrum of a single scalar variable, such as the temporal 
component of the metric perturbation in the longitudinal gauge \cite{Mukhanov}, often denoted by $\Phi$. This variable plays the role of the Newtonian potential in the non-relativistic limit. In Fourier space the gravitational potential 
is expressed as  
\begin{equation}
\Phi(\vec x, \eta)=(2\pi)^{-3/2}\int \Phi_{\vec k}(\eta)e^{i \vec k\cdot\vec x}\,d^3k. \label{fourierphi}
\end{equation}
For perturbations with $k\eta_{eq}\gg1$, entering the horizon well before equality($\eta\ll\eta_{eq}$), radiation dominates over dark matter and baryons. Neglecting the decaying mode on supercurvature scales, the time dependence of 
such modes for $\eta \ll \eta_{eq}$ is then given by \cite{Mukhanov}
\begin{equation}
\Phi_{\vec{k}}(\eta) \approx \Phi^{0}_{\vec{k}}\ \left[G(k\eta/\sqrt{3})+ \kappa s(\eta) H(k\eta/\sqrt{3})\right], \label{phik}
\end{equation}
where $\kappa = \Omega_{DM}/\Omega_{M}\approx 0.84$ is the fraction of non-relativistic matter in the form of dark matter, and the time dependence is given in terms of 
\begin{equation}
G(x) \equiv {3\over x^{2}}\Big[\frac{\sin x}{x}-\cos x\Big], \quad\quad H(x) \equiv  {9\over 2 x^2}\left[{\bf C} -{1\over 2} + \ln x\right] \Theta(x-1).\label{functionG}
\end{equation}
Here, ${\bf C}\approx 0.577$. The initial amplitudes $\Phi_{\vec k}^0$ of the gravitational potential on superhorizon scales are Gaussian distributed, with variance given by 
\begin{equation}
\langle \Phi_{\vec k}^0 \Phi_{\vec k'}^0\rangle =
\sigma^2_\Phi(k) \delta^{(3)}(\vec k+\vec k'). \label{variance}
\end{equation}
The functions $G$ and $H$ represent the contribution of radiation and matter density perturbations, respectively. Which one dominates the torque will depend on the time $\eta_b$ when binaries decouple from the Hubble flow, which in turn is related to the co-moving binary size [see Eq. (\ref{sb})]. At early times, matter is subdominant, and the contribution of matter perturbations to $\Phi$ is suppressed by the scale factor 
$s=(\eta/\eta_{eq})$ in front of $H$ in Eq. (\ref{functionG}). 
The factor $\Theta(x-1)$ in $H$ should not be taken too literally, it is just meant to indicate that the expression is only valid after the modes cross the horizon, $k\eta_{eq} \gg 1$, and the logarithmic growth begins.

The angular momentum per unit reduced mass of the binary, $\vec{\ell}$, can be written as a time integral of the tidal torque exerted by the gravitational potential. Denoting by $\vec x_1(\eta)$ amd $\vec x_2(\eta)$ the co-moving positions of the two members of the binary, we have:
\begin{equation}
\vec{\ell}
=-\int{\vec x(\eta)\times[\vec\nabla\Phi(\vec x_2,\eta)-\vec\nabla\Phi(\vec x_1,\eta)]}\,s(\eta)d\eta, \label{angular}
\end{equation}
where $\vec x(\eta) = \vec x_2-\vec x_1$ is the relative co-moving coordinate, which is time dependent from the time $\eta_b$ when the binary decouples from the Hubble flow [see Eq. (\ref{timedependence})]. 

In order to calculate the variance of the angular momentum, we will work at lowest order in the gravitational potential, so that inside the integrand in Eq. (\ref{angular}) we can use the unperturbed head-on trajectory, which we shall take along the $z$ axis, with relative coordinate:
\begin{equation}
\vec x = x \hat e_z.
\end{equation}
Assuming that the center of mass is at the origin of coordinates, the positions of the two PBHs are given by $\vec x_2 = (m_1/M)\vec x$ and $\vec x_1= -(m_2/M)\vec x$. Using (\ref{variance}) and (\ref{angular}) we have
\begin{equation}
\langle \ell^i \ell^j \rangle =  {\epsilon^{zmi}\epsilon^{znj} \over 2\pi^3} \int \sigma^2_{\Phi}(k) k_m k_n  F^*F d^3\vec k ,
\end{equation}
where the indices $i,j,m,n$ can only take values $x$ or $y$ and
\begin{equation}
F={1\over 2}\int_0^{\eta_{eq}} d\eta s(\eta) x [G(k\eta/\sqrt{3})+\kappa s(\eta)H(k\eta/\sqrt{3})] \left(e^{ik_z x{m_1\over M}} - e^{-ik_z x{m_2\over M}}\right). \label{fintegral}
\end{equation}
Introducing spherical coordinates in momentum space,  $k_x=k \sin\theta\cos\phi$, $k_y = k \sin\theta\sin\phi$, $k_z= k \cos\theta $, and integrating over $\phi$, we have
\begin{equation}
\langle \ell^2 \rangle= {1 \over\pi^2} \int dk k^4 \sigma^2_{\Phi}(k)\int_{-1}^1 dw (1-w^2) |F(k,x_0,w)|^2, \label{variancell}
\end{equation}
where we have used $\langle \ell^2 \rangle=\langle \ell^i \ell^j \rangle \delta^{ij}= 2\langle \ell^x \ell^x \rangle$ and we have introduced the change of variable $w=\cos\theta$. 

To estimate the integral $F$ we first note that the radiation $G$, and matter $H$ terms make their contribution at very different times. Consider a binary with initial separation $x_0$ that decouples from the Hubble flow at the conformal time $\eta_b$.  From (\ref{a}) and (\ref{semimajor}), these two scales are widely separated, and parametrically related by 
\begin{equation}
k_0^{-1} = x_0 \sim 5.4\cdot 10^{-6} m^{1/3}\lambda^{-2/3} \eta_b \ll \eta_b. \label{x0andetab}
\end{equation}
In general, all perturbations make their contribution to the torque at times $\eta \lesssim \eta_b$. After that, the binary starts oscillating, its co-moving size shrinks, and tidal gradients decay in inverse proportion to the scale factor.
Perturbations in radiation start oscillating once they enter the horizon, and make most of their contribution to the torque at $\eta\sim k^{-1}\lesssim \eta_b$, while matter perturbations make their contribution 
near the time $\eta\sim \eta_b$, regardless of their wavelength.

\subsection{Perturbations in radiation}

Let us start by considering small binaries, for which $\lambda \ln(k_0\eta_b) \ll 1$. These form deep in the radiation dominated era, at the time when $s=s_b(\lambda)\sim \lambda \ll 1/\ln(k_0\eta)$. In this case, perturbations in radiation dominate over matter perturbations. From Eq. (\ref{x0andetab}), and assuming masses in the stellar range, the logarithm is of order 10, and so this condition requires $\lambda \ll 0.1$.
In this regime, we can neglect $s(\eta)H(k\eta/3)$ relative to $G(k\eta/3)$ for $\eta\lesssim \eta_b$, and we have
\begin{equation} 
F\approx {x_0\over 2}\left(e^{ik w x_0{m_1\over M}}-e^{-ikw x_0{m_2\over M}}\right) \int_0^{\eta_{eq}}  d\eta\, s\, G\left({k\eta\over\sqrt{3}}\right). \quad(k\eta_b \gtrsim 1)
\end{equation}
Due to the oscillating nature of $G$ for $\eta \gg k^{-1}$, the integral is dominated by early times $\eta\ll k^{-1} \lesssim \eta_b$, where $x=x_0 \chi(\eta)/s$ is approximately constant, $x\approx x_0$, and can be taken out of the integral (\ref{fintegral}). Noting that 
\begin{equation}
 \int_0^{\eta_{b}}  d\eta s G\left({k\eta\over\sqrt{3}}\right) = {9\over k^2 \eta_{eq}}\left[ 1- {\rm sinc}\left({k\eta_{b}\over\sqrt{3}}\right)\right],
\end{equation}
it is clear that the contribution of modes outside the horizon at the time of binary formation, $k\eta_b\ll 1$, is suppressed. For modes with $k\eta_b \gg 1$ we can approximate
\begin{equation}
|F|^2 \approx {81 x_0^2\over k^4\eta^2_{eq}} \sin^2\left({k w x_0\over 2}\right). \label{Fapprox}
\end{equation}
Substituting in (\ref{variancell}), and using
\begin{equation}
\int_{-1}^1 dw (1-w^2) \sin^2\left({k w x_0\over 2}\right) = {2\over 3}[1-G(k x_0)],
\end{equation} 
where the function $G$ is defined in (\ref{functionG}), we have\footnote{In determining the numerical coefficient in front of (\ref{variancell2}), we have used, from  (\ref{barx}) and (\ref{a}),
\begin{equation}
GMa =G\beta \lambda \bar m x = G\beta \bar m {x_0^4 \over f \bar x^3} = \beta x_0^4 {2\pi G\over 3}\rho_{eq}={\beta\over 4}x_0^4 H^2_{eq}\approx {x_0^4 H^2_{eq}\over 20}.
\end{equation}
Also, we have used $H_{eq}^2\eta_{eq}^2 \approx 1$. Note that we are using the convention where the scale factor is equal to unity at the time of equality.}
\begin{equation}
\sigma^2_{cp(rad)}\equiv \langle j_{cp}^2 \rangle = 
{\langle \ell^2 \rangle \over GMa} \approx 2.2 \cdot 10^3 \int {dk\over k}\ P_{\Phi}(k)\left[{1-G(k x_0)\over (k x_0)^2}\right]\left[ 1- {\rm sinc}\left({k\eta_{b}\over\sqrt{3}}\right)\right]^2. \label{variancell2}
\end{equation}
The subindex in $j_{cp}$ indicates that this is due to cosmological perturbations, as opposed to the neighboring black holes which we considered in the previous Section. 
Here we have introduced the standard expression for the primordial power spectrum 
\begin{equation}
P_{\Phi}(k) = {\sigma^2_\Phi k^3\over 2\pi^2}.
\end{equation}
In standard slow-roll inflationary scenarios, $P_{\Phi}(k)$ is nearly independent of $k$. In that case, 
the two factors in square brackets in Eq. (\ref{variancell2}) play the role of the infrared and ultraviolet cut-off which regulate the logarithmic behaviour of the integral. Note that 
\begin{equation}\label{gapprox}
{1-G(k x_0)\over (k x_0)^2} \approx
\begin{cases}
{1\over 10}, \quad (k x_0 \ll 1) \\
{1\over (k x_0)^2}. \quad (k x_0 \gg 1)
\end{cases}
\end{equation}
while 
\begin{equation}
1- {\rm sinc}\left({k\eta_{b}\over\sqrt{3}}\right) \approx
\begin{cases}
{(k\eta_b)^2\over 18}, \quad (k \eta_b \ll \sqrt{3}) \\
{1}. \quad (k \eta_b \gg \sqrt{3})
\end{cases}
\end{equation}
Therefore, for a nearly flat power spectrum, the integral will be dominated by the range $\eta_b^{-1}\lesssim k \lesssim k_0 = x_0^{-1}$,
\begin{equation}
\langle j^2_{cp}\rangle \approx  2.2\cdot 10^2 \int_{\sqrt{3}\eta_b^{-1}}^{k_0}{dk\over k}P_{\Phi} \approx 2.2\cdot 10^2 \ln\left({k_0\eta_b\over\sqrt{3}}\right) P_\Phi, 
\end{equation}
where
\begin{equation}
\ln\left({k_0 \eta_b\over \sqrt{3}}\right) \approx 9.6 + \ln\left[\left({\lambda\over 0.3}\right)^{2/3} \left({m\over 30}\right)^{-1/3}\right] . \label{L0}
\end{equation}
In the last approximate equality, we have used  (\ref{x0andetab}).

It may seem counterintuitive that the dispersion in $j$ receives contributions from a wide range of scales, since at the time $\eta_b$ when the binary forms, the amplitude of the gravitational potential for modes within the horizon falls off as $k^{-2}$. However, the torque at wavelenths larger than $x_0$ depends on second derivatives of the potential, which brings in a factor of $k^2$. As a result, the contribution is independent of scale in the range we  are considering.  Physically, the effect takes place well before the binary decouples from the Hubble flow, at the time $\eta\lesssim k^{-1}$. Hence, it seems appropriate to refer to this as the contribution of the peculiar velocities of the PBHs to the orbital angular momentum at the time when the binary forms.

It is also worth noting that the prefactor in front of the logarithm is independent of the parameters characterizing the binary. Each decade in wavelength gives the same contribution to $\langle j^2\rangle$, and the dependence on parameters such as massess and semi-major axis, is only through the range of scales contributing to the logarithm. This is in contrast with the contribution from matter perturbations, which we now review.

\subsection{Adding matter perturbations}

Matter perturbations can be included along similar lines. One difference is that their effect on the binary occurs near the time $\sim \eta_b$ when the binary starts oscillating, and we cannot ignore the time dependence of the separation $x$ in the integral (\ref{fintegral}) which gives $F(k,x_0,w)$. For $k x_0 \gg 1$, this has a dependence on $m_1$ and $m_2$ which, unlike the radiation case, is hard to disentangle in general.

For $k x_0\ll 1$, we may use the approximation 
\begin{equation}
e^{ikwx{m_1\over M}} - e^{-i k w x{m_2\over M}} \approx i k wx,
\end{equation}
in the integral (\ref{fintegral}). With this approximation, the dependence on masses disappears and we obtain the total contribution of radiation and matter perturbations as \footnote{In order to obtain the numerical coefficient in front of the matter contribution, we have used \cite{Kam,RSVV}
$\int_0^1 ds (\chi^2/s^2) \approx 0.3 \lambda (2\bar m/M)$ to do the intergral of the second term in (\ref{fintegral}). Also, we have used $\kappa\approx 0.84$ for the ratio of dark matter density to the total non-relativistic matter density, and we have ignored the slow logarithmic dependence in $\eta$. Since the 
integral is dominated by $\eta\sim\eta_b$, we have used the value $\eta=\eta_b$ inside the logarithm in the mode function $H$.}
\begin{equation}
F \approx i kw {9 x_0^2 \over k^2\eta_{eq}} \left\{ 1 + 0.38 \lambda \left({2\bar m\over M}\right) [L_0 + \ln(kx_0)]\right\}, \label{ftotal}
\end{equation}
where 
\begin{equation}
L_0\equiv \ln(k_0\eta_b/\sqrt{3}) +{\bf C}-(1/2) \approx 9.7. \label{L02}
\end{equation}
Here, we have used (\ref{L0}), neglecting the small logarithmic dependence in $\lambda$ and $m$.


Substituting (\ref{ftotal}) in (\ref{variancell}) and performing the $w$ integration, we immediately find the total variance of the orbital parameter due to long wavelength cosmological perturbations:
\begin{equation}
\sigma^2_{cp}\equiv \langle j^2_{cp}\rangle \approx  2.2\cdot 10^2 \int_{\sqrt{3}\eta_b^{-1}}^{k_0}{dk\over k}   \left[ 1 + 0.38 \lambda \left({2\bar m\over M}\right) [L_0 + \ln(kx_0)]\right]^2 P_{\Phi}(k).\label{totvar}
\end{equation} 
The first term in the square brackets in (\ref{ftotal}) corresponds to radiation, while the second one, accompanied by the factor of $\lambda$, corresponds to matter perturbations. The latter become subdominant for sufficiently 
small $\lambda \lesssim 0.27$.  In view of our earlier discussion in Subsection \ref{ssthird}, for $f\gg 3\cdot 10^{-3}$ the rates are dominated by small binaries, with $\lambda\lesssim \lambda_*\ll 1$. Hence, it appears that 
perturbations in radiation may be as relevant for observations as the matter perturbations which have been considered in earlier analysis. 


Let us now turn to a discussion of the effect of such perturbations on the merger rates.

\begin{figure}
\begin{center}
\includegraphics[scale=0.75]{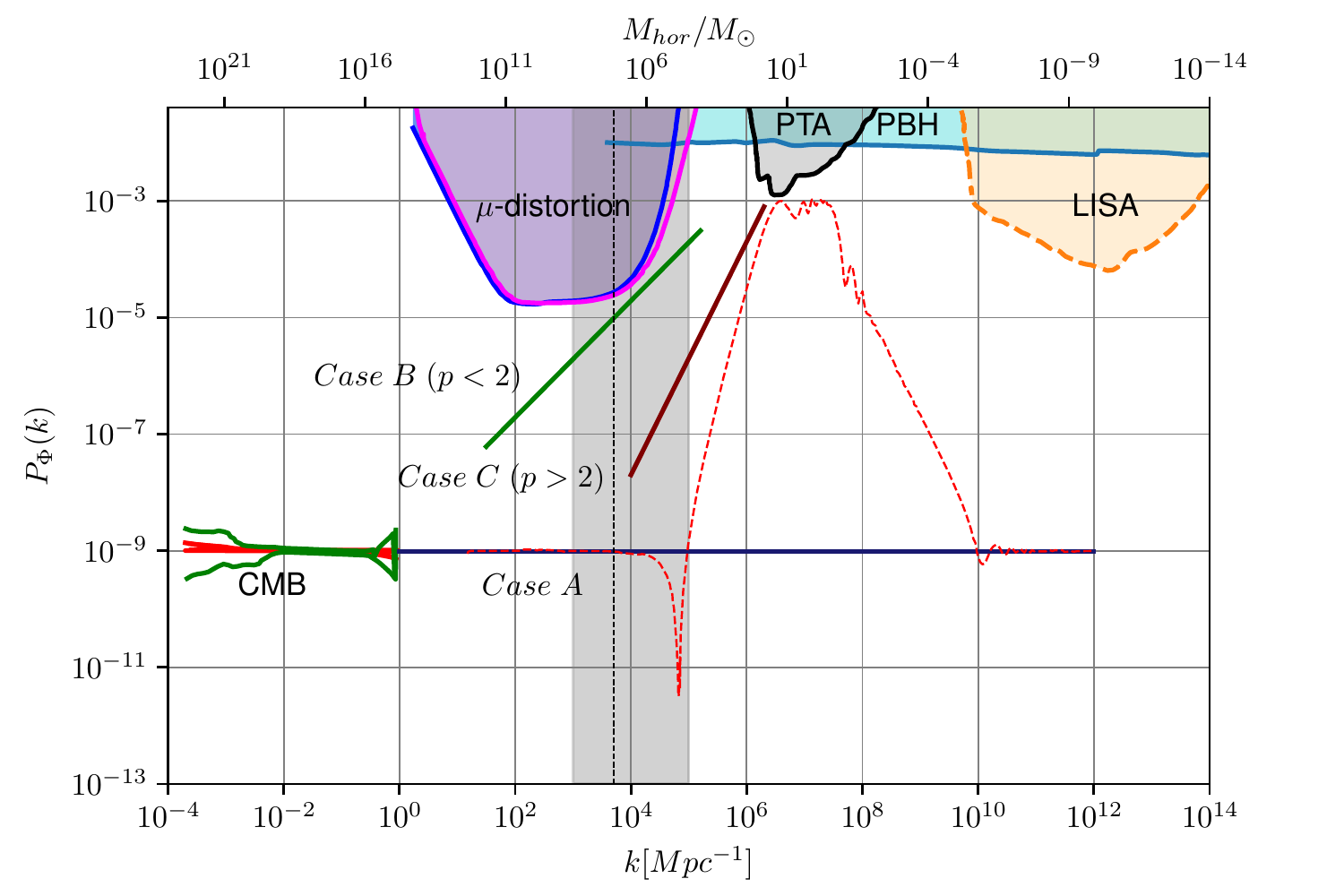}
\caption{Current bounds on the power spectrum $P_\Phi$ of cosmological perturbations
(see e.g. \cite{current,BCP}). Aside from the constraints from $\mu-$distorsions of the CMB, we also display the pulsar timing array (PTA) and LISA bounds, which constrain the production of gravitational waves from scalar perturbations at second order in perturbation theory. The PBH bound near the top of the figure is intended to represent the level at which the probability of PBH formation would high enough to be in conflict with current constraints on their cosmic abundance. We consider the effect of cosmological perturbations on the merger rate of PBH binaries in three different cases, labeled as case A,  B and C, consistent with the observational constraints. The vertical shaded band corresponds to the scales $k_0$ characteristic of binaries which would be merging at the present time, in the mass range $m\sim 1-100$. The vertical dashed line corresponds to the pivot scale $k_B \approx 5\cdot 10^3 Mpc^{-1}$ in Eq. (\ref{powerB}). For illustration, in dashed red line we plot the power spectrum of an inflationary model, which raises steeply with spectral index $p\approx 4$ up to a scale $k_C\approx 3\cdot 10^6 Mpc^{-1}$, leading to a rather broad peak spanning one order of magnitude or so. This corresponds to a model \cite{BCP} where the inflaton goes from slow roll to fast roll and then back to slow roll, through a discrete sequence of values of the second slow roll parameter $\eta$ (see Fig. 10 of \cite{BCP}).}
\label{power}
\end{center}
\end{figure}

\section{Effect of perturbations on the merger rates}

\label{mergerrates}

The size of binaries at the time of formation (shaded in gray in Fig. \ref{power}) is at intermediate co-moving scales which are much smaller than those probed by CMB temperature anisotropies or large scale structure. Indeed, from Eqs. (\ref{a}) and (\ref{semimajor}), we have
\begin{equation}
k_0 = {1\over x_0} \approx 0.5 \cdot 10^6 (m \lambda)^{-1/3} k_{eq}. \label{k0}
\end{equation}
Taking into account that $\lambda$ is in the range (\ref{lambdarange}), for stellar mass black holes $m \sim 1-100$ we have $k_0 \sim (10^5-10^7) k_{eq}$, which corresponds to the present co-moving scale in the range\footnote{Throughout this paper, we adopt the convention that the scale factor $s$ is equal to $1$ at the time of equality. Thus, to avoid confusion, we will refer to the present day wave number by $\bar k = z_{eq} k$. Note that relations such as Eq.(\ref{k0}) between $k_0$ and $k_{eq}$ are valid in both conventions, since the factor of $z_{eq}$ applies to both sides of the equation.}  $\bar k \sim (10^3- 10^5) Mpc^{-1}$.

The angular momentum of PBH binaries may be affected by the power spectrum $P_\Phi$ of cosmological perturbations over a very wide range of scales. As illustrated  
in Fig. \ref{power}, such power spectrum is poorly constrained on scales smaller than $3 Mpc^{-1}$, and here we would like to explore the consequences this uncharted territory
might have on the merger rate of binaries.\footnote{In Fig. \ref{power}, we are ignoring bounds which are related to the abundance of ultra-compact mini-halos. Such constraints depend on the nature of dark matter, and could be absent in certain models (e.g. if dark matter belongs to a hidden sector). For a recent discussion, see \cite{current} and references therein.}
 For this purpose, let us consider three distinct behaviours which may capture the generic effect of an enhanced power spectrum at small scales. 
These are labeled case A, B and C in Fig. \ref{power}. Let us consider them in turn.

\subsection{Case A: Nearly scale invariant cosmological perturbations}
\label{flatplat}

Consider a nearly scale invariant power spectrum of the form
\begin{equation}
P_\Phi \approx A_\Phi \left({k\over k_*}\right)^{n_s-1}. \label{standard}
\end{equation}
This is consistent with observations of the CMB and large scale structure on cosmological scales, down to $\bar k \sim 3 Mpc^{-1}$, with $n_s\approx .97$, $A_\Phi \approx .97\cdot 10^{-9}$ and $\bar k_*\approx 0.05 Mpc^{-1}$ \cite{Akrami}.
A minimal assumption we can make, consistent with standard slow roll inflationary models, is that the nearly flat spectrum 
can be extrapolated down to the co-moving size of binaries. Since the tilt is rather small, we may approximate $P_\Phi \approx const.$ over the range of interest.

\begin{figure}
\begin{center}
\includegraphics[scale=0.7]{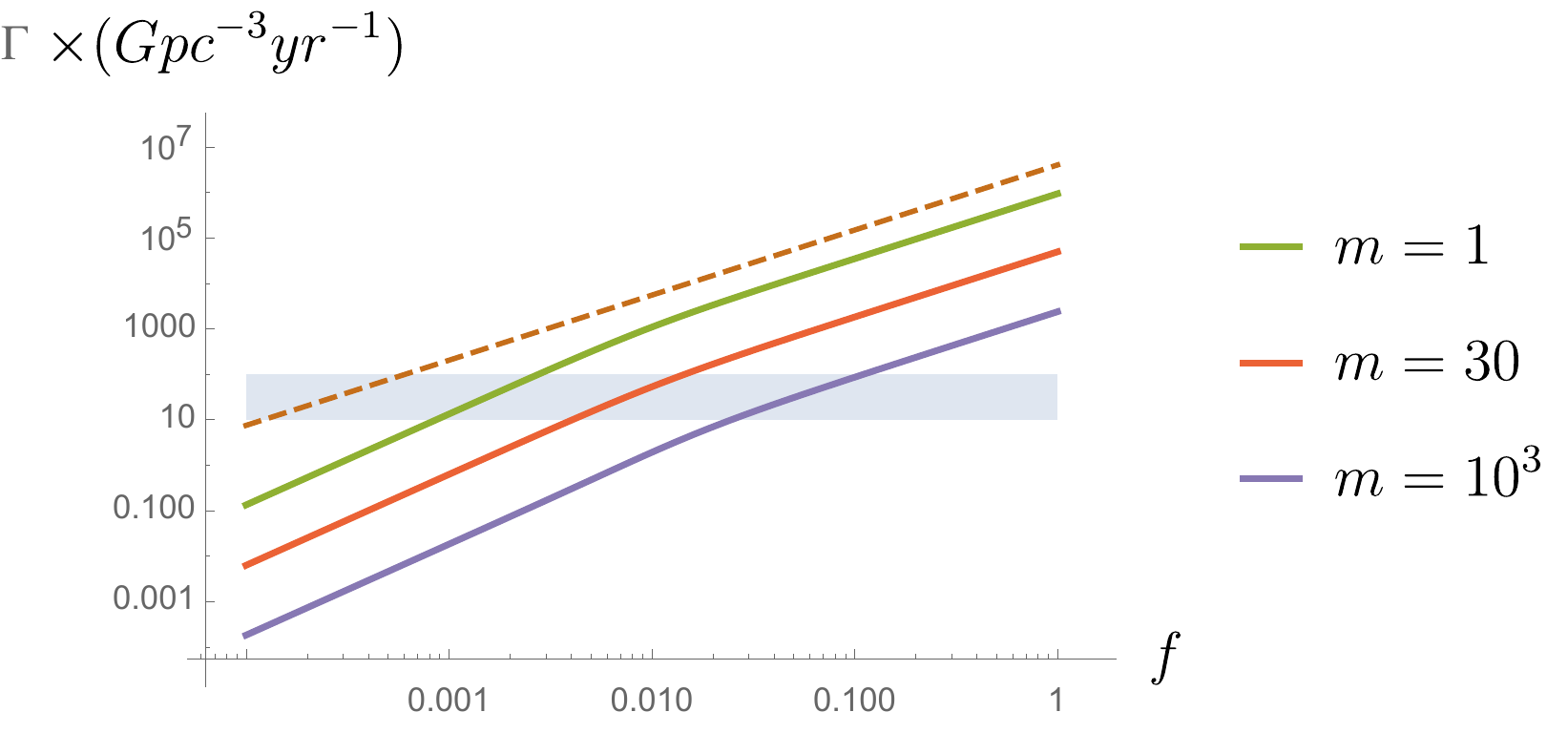}
\caption{Here we plot the merger rate as a function of $f$ (thick lines) for different values of $m$, assuming  a scale invariant spectrum of cosmological perturbations with amplitude $P_\Phi \sim 10^{-9}$ (Case A). At small $f$, cosmological perturbations control the angular momentum of binaries merging today. The knee at $f\sim 10^{-2}$ arises because at higher $f$ the effect of neighboring PBHs becomes dominant. Here we have used $Y_{min}=2$ for the infall radius. 
For comparison, we include the dotted line, corresponding to the analytic estimate (\ref{rateflarge}) which ignores cosmological perturbations and binary infalls (here we use $m=1$).}
\label{caseA}
\end{center}
\end{figure}

Using (\ref{totvar}) we then find
\begin{equation} 
\sigma^2_{cp} =\langle j_{cp}^2\rangle \approx 2.1 \cdot 10^{3} \left[1+ 3.7 \lambda \left({2\bar m \over M}\right) + 4.5 \lambda^2 \left({2\bar m\over M}\right)^2\right]P_\Phi, \label{dispplateau}
\end{equation}
Assuming the CMB normalization $P_\Phi \approx 10^{-9}$, we have $\sigma^2_{cp} \sim (2-9) 10^{-6}$.
Hence, for moderate values of $\lambda$, cosmological perturbations may easily provide an angular momentum comparable to $j_0$, given in (\ref{j0}), necessary for binaries to have a life-time comparable to the age of the universe. 
Still, neighboring PBH contribute to $j$ with variance given by (\ref{sigmanb}) \cite{RSVV},
\begin{equation}
\sigma^2_{nb} \approx {K \over 4Y_{min}} f^2 \lambda^2\left({2\bar m\over M}\right)^2, \quad (Y_{min}\gtrsim 2), \label{varnb}
\end{equation}
where we are excluding binaries which may be disrupted by a PBH at distances smaller than $Y_{min}$. Thus, at low $f$ the effect of neighboring PBHs on the angular momentum is negligible compared to that of cosmological perturbations, while the latter effect can become important only at higher $f$. This is illustrated in Fig. \ref{caseA}.


To better understand the relative importance of the two effects, it is illustrative to look at the differential merger rate in the integrand in Eq.  (\ref{intrate}):
\begin{equation}
W(\lambda)= 2 {j_0^2 \over \sigma^2} e^{-{j_0^2\over \sigma^2}} e^{-Y{min}}, \label{wlambda}
\end{equation}
where $\sigma^2 = \sigma^2_{nb} + \sigma^2_{cp}.$ 

\begin{figure}
\includegraphics[scale=0.5]{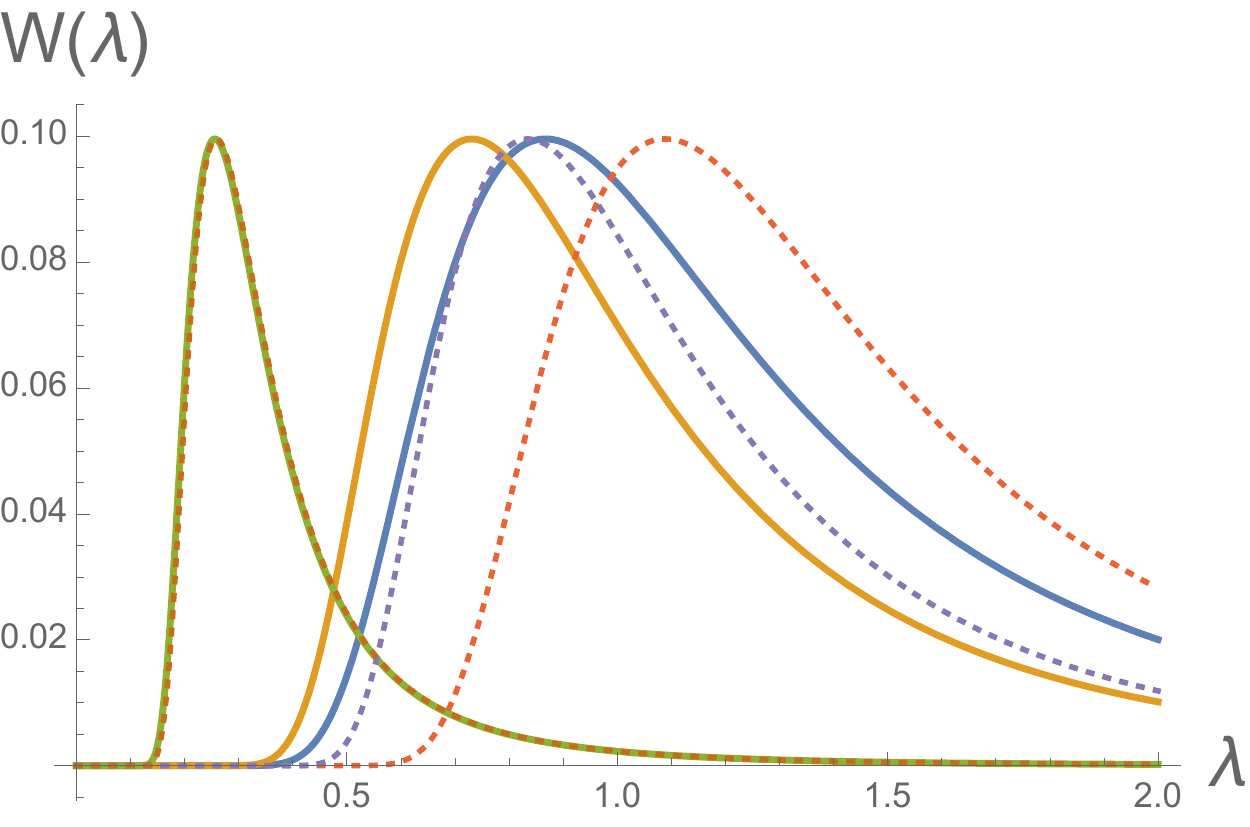}
\includegraphics[scale=0.5]{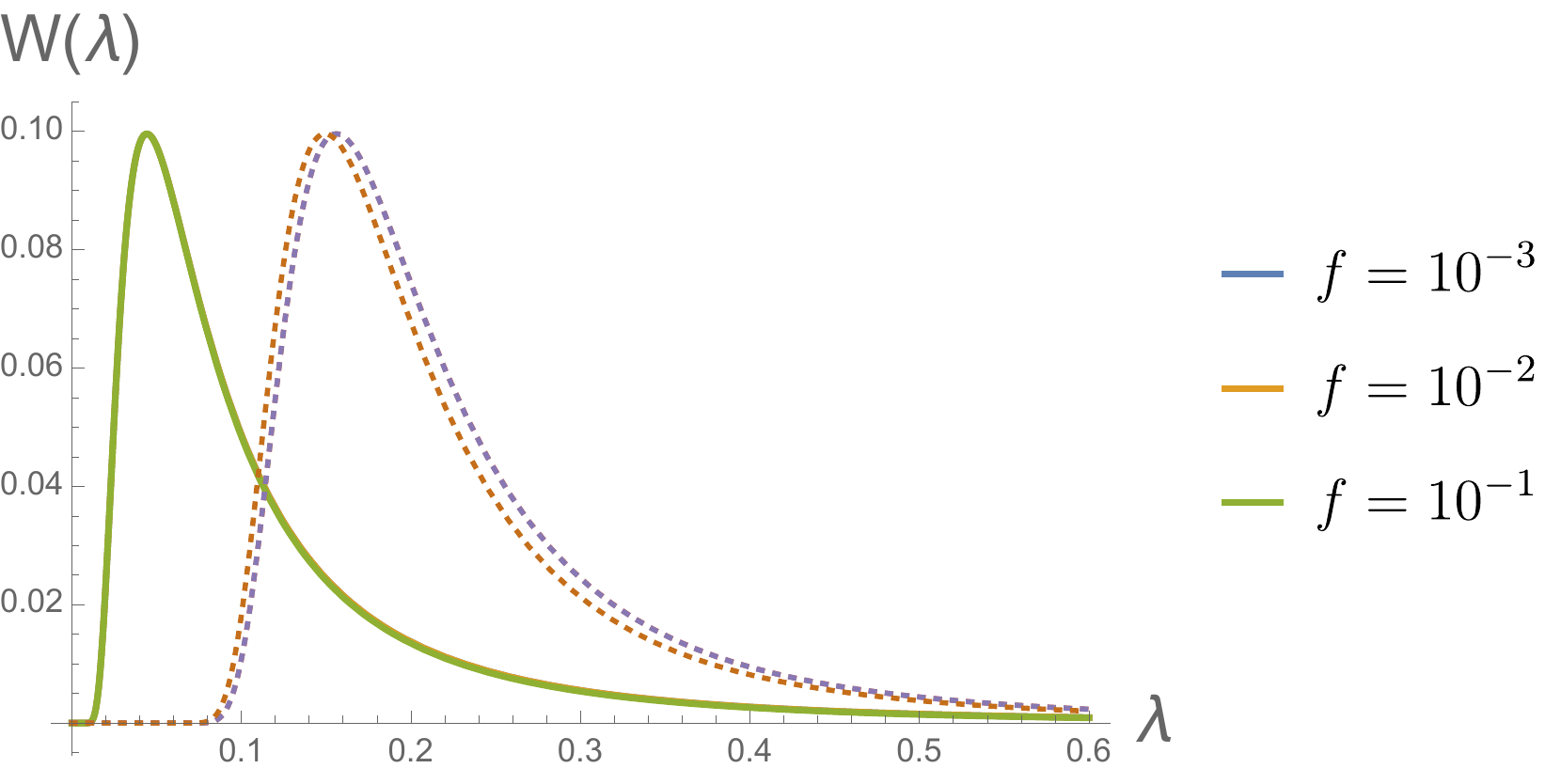}
\caption{The left pannel shows the differential rate $W$ in the integrand of Eq. (\ref{intrate}), for $m=30$ and different values of $f$, in the case of a scale invariant spectrum of cosmological perturbations with amplitude $P_\Phi = 10^{-9}$ (Case A). The curves move from right to left with increasing values of $f$. At low $f\lesssim 10^{-2}$,  cosmological perturbations give the dominant contribution to $j$,, while for higher $f$, cosmological perturbations are subdominant compared with the effect of neighboring PBHs. The thick curves include both matter and radiation perturbations, while perturbations in radiation are ignored in the dotted curves. We see that perturbations in radiation cause a shift of the peak towards smaller $\lambda$, which means that the binaries which are merging today decouple from the Hubble flow somewhat earlier. The right pannel corresponds to an enhanced power spectrum, which is nearly scale invariant at intermediate scales (Case B). In this case, cosmological perturbations provide the dominant contribution to the angular momentum of binaries which are merging today, for all values of $f$. The differential rate $W$ is plotted for a power spectrum amplitude $B_\Phi=10^6$ and three different values of $f$ in thick lines. The three cases are degenerate, which implies that cosmological perturbations, and not the neighboring PBHs, provide the dominant torque. Note that $W$ peaks at $\lambda \sim 0.05$. This corresponds to binaries which decouple from the Hubble flow deep in the radiation dominated era, when perturbations in non-relativistic matter are small relative to the perturbations in the radiation fluid. This is confirmed by the curves in dotted lines, where perturbations in radiation have been ignored, and which are quantitatively different (the curves with $f\lesssim 10^{-2}$ are degenerate in this case). Dominance of radiation perturbations grows even stronger at higher values of $B_{\Phi}$.}
\label{wcaseA}
\end{figure}

The function $W(\lambda)$ for Case A is plotted in thick lines in the left pannel of Fig. \ref{wcaseA}, for different values of $f$, and $m=30$. At low $f\lesssim 10^{-2}$, the effect of neighboring PBHs is negligible. This corresponds to the values below the knee in Fig. \ref{caseA}. In this case Fig. \ref{wcaseA} shows that for the low amplitude cosmological plateau with $P_\Phi \approx 10^{-9}$ the dominant contribution to $W$ is at $\lambda\sim 1$.  This value corresponds to a time of binary formation $\eta_{b}\sim \eta_{eq}$. Even at this relatively late time the effect of relativistic matter perturbations has a noticeable effect on the position of the peaks. For comparison, we plot in dotted lines the case where perturbations in radiation are ignored. At higher $f\sim 10^{-1}$  we see from Fig. \ref{wcaseA} that the rate is dominated by lower $\lambda \sim 0.1$. For such values of $\lambda$, binaries decouple earlier in time, when perturbations in radiation would be more important than those in non-relativistic matter. But in fact, both are negligible, since for such low values of $f$, the variance in $j$ is dominated instead by the torque from neighboring PBHs. The situation is different when we consider an enhanced power spectrum (Case B), which is plotted in the right pannel of Fig. \ref{wcaseA}. 

Before moving on to a discussion of enhanced power spectra, let us consider the parameter $\alpha$ in the presence of the standard scale invariant spectrum of cosmological perturbations (\ref{standard}). Note that the variance (\ref{varnb}) dominates at high $f$, whereas the last term in (\ref{dispplateau}) dominates at low $f$. Both terms have exactly the same dependence on $\lambda$ and on the masses. They only differ in the explicit dependence of the variance on $Y_{min}$, which is absent at low $f$. Hence, by the same argument which lead us to Eq. (\ref{alphamany}), it is straightforward to conclude that 
\begin{equation}
{36\over 37} < \alpha < {36 \over 37} + {21\over 74}, \label{alphaA}
\end{equation}
where the lower and upper bounds correspond to the limiting behaviours for low and high $f$ respectively. 

\subsection{Case B: Enhancement of the power spectrum at intermediate scales.}  \label{intermediate}

Bounds on the amplitude of the power spectrum above $\bar k \sim 10  Mpc^{-1}$ are rather loose. The absence of CMB spectral distortions gives an upper bound at intermediate scales in the range $\bar k \sim 10-10^5 Mpc^{-1}$ \cite{Chluba,current} 
\begin{equation}
P_\Phi \lesssim 10^{-5}, \label{smoot}
\end{equation}
several orders of magnitude higher than the nearly scale invariant plateau observed on cosmological scales. 

Hence, let us consider a generic power spectrum of the form
\begin{equation}
P_\Phi(k) = B_\Phi \left({k\over k_B}\right)^p,  \quad (0 \leq p< 2) 
\label{powerB}
\end{equation}
where for convenience we choose the pivot scale to be $k_B= 5 \cdot 10^5 k_{eq}$, corresponding to the present co-moving scale of $5 \cdot 10^3 Mpc^{-1}$. From Eq. (\ref{k0}), this is by order of magnitude the co-moving size of the binaries, and we see from Fig. \ref{power} that the power spectrum can be rather high at those scales, allowing for an enhanced amplitude up to $B_\Phi \lesssim 10^{-5}$. Let us now discuss the impact of this enhancement as a function of $p$. Here we focus on $p<2$, leaving the discussion of steeper power spectra for the next Subsection.

\begin{figure}
\begin{center}
\includegraphics[scale=0.7]{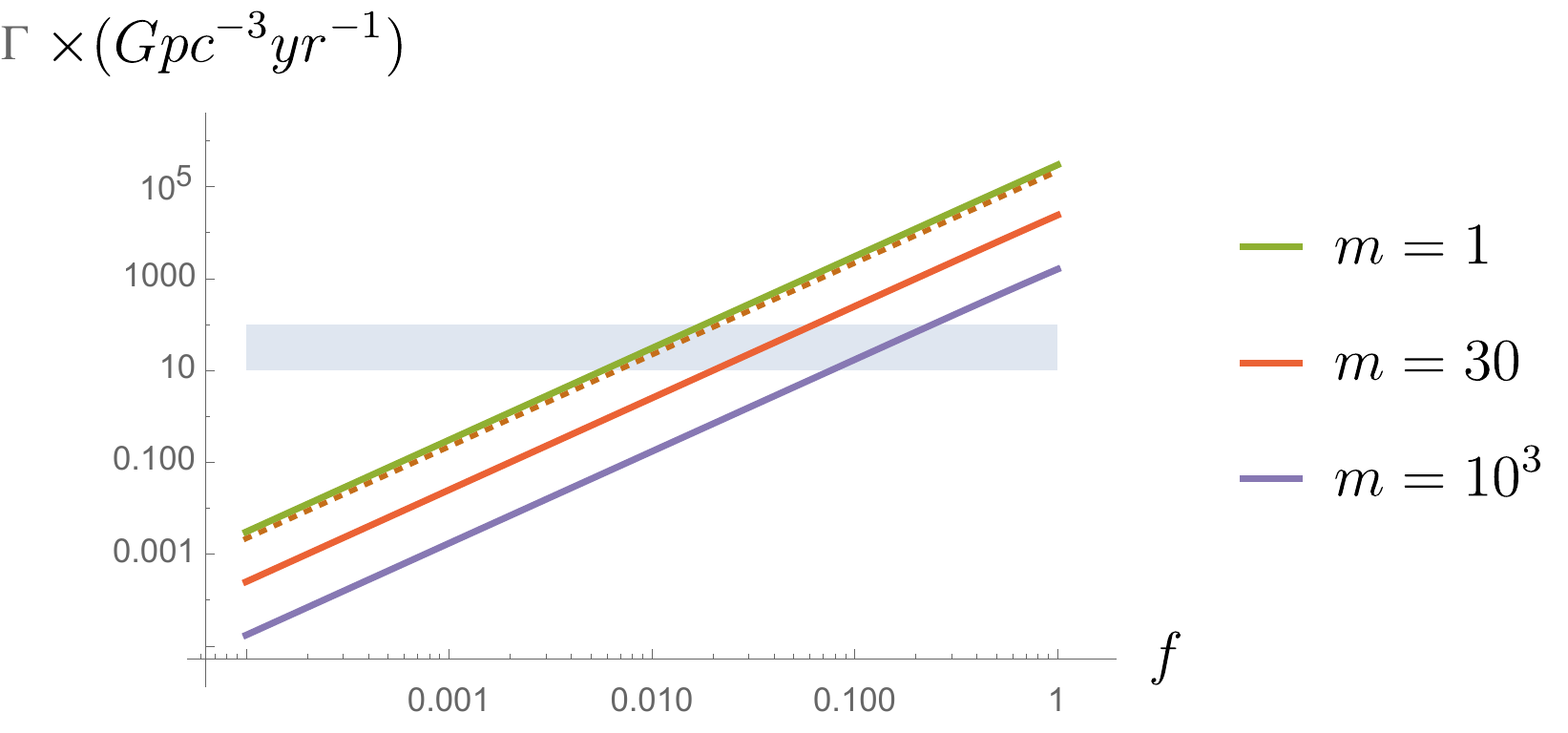}
\caption{The thick lines represent the rates corresponding to a nearly scale invariant spectrum which is enhanced to the saturate the observational bound, $P_\Phi \lesssim 10^{-5}$, at the intermediate scales $10-10^5 Mpc^{-1}$. This corresponds to Case B with $p\approx 0$, discussed in Subsection \ref{intermediate}, and we have used $Y_{min}=2$. The dotted line shows the analytic expression (\ref{aapprox}) for $m=1$.} 
\label{rateB}
\end{center}
\end{figure}

The case $p=0$ is special, since the variance $\sigma^2_{cp}$ depends logarithmically on the width of the plateau in the power spectrum. For illustration, let us consider a scale invariant spectrum of amplitude $B_\Phi$ between the scale $k_{min} = 10^3 k_{eq}$ and $k_{max}\gtrsim 10^7 k_{eq}$. Then, using  (\ref{totvar}), we find\footnote{In this estimate, we neglect a small subleading dependence in $\lambda$ and $m$, assuming $|\ln\left[(\lambda/0.3)(m/30)\right]|\ll 16$.}
\begin{equation}
\sigma^{2\ (p=0)}_{cp} =\langle j_{cp}^2\rangle \approx 1.2 \cdot 10^{3} \left[1+ 5.2 \lambda \left({2\bar m \over M}\right) + 7.5 \lambda^2 \left({2\bar m\over M}\right)^2\right] B_\Phi. \label{dispplateauB}
\end{equation}
Comparing with (\ref{varnb}), it is clear that for $f\lesssim 0.1$ cosmological perturbations will dominate over the effect of neighboring PBH for amplitudes $B_\phi \gtrsim 10^{-7}$, which are well within the range allowed by observational constraints. This is illustrated in the right pannel in Fig. \ref{wcaseA}, where we see that the rate is dominated by $\lambda\lesssim 0.1$ Hence, we may approximate
\begin{equation}
\sigma^{2\ (p=0)}_{cp} \approx 1.2 \cdot 10^{3} B_\Phi. \quad (B_\Phi\gtrsim 10^{-7}) \label{dispplateauBapprox}
\end{equation}
Using (\ref{j0}) we then have
\begin{equation}
{j_0^2\over \sigma^2}  \approx \left({\lambda\over \lambda_B^{(p=0)}}\right)^{-32/21},
\end{equation}
where 
\begin{equation} 
\lambda_B^{(p=0)} = \left[{2.1 \cdot 10^{-9} \over B_\Phi} \left({M\over 2 \bar m}\right)^2 \left({ 4 m_1 m_2 \over M^2}\right)^{2/7} \right]^{21/32} m^{10/32}.
\end{equation}
For sufficiently high $B_\phi \gtrsim 10^{-7}$ we have $\lambda_B^{(p=0)} \ll 1$, and we can ignore the upper limit of integration in (\ref{intrate}). Then we have:
\begin{equation}
\Gamma_M(t_0) \approx {\rho_m \over M_\odot t_0} {f^2 \over 14 m} {21\over 32}\Gamma\left({11\over 32}\right) \lambda_B^{(p=0)} e^{-Y_{min}},
\end{equation}
and using (\ref{numerics}) and (\ref{dispplateauBapprox}) we have
\begin{equation}
\Gamma_M(t_0) \approx {7.8\cdot 10^4 f^2 \over Gpc^3 Yr} {m^{-11/16}\over \sigma^{21/16}}  \left({M\over 2 \bar m}\right)^{21\over 16} \left({ 4 m_1 m_2 \over M^2}\right)^{3/16} e^{-Y_{min}}.\label{onemore}
\end{equation}
We note that the same parametric dependence is obtained if we simply assume that all binaries are born equal, with the same initial dimensionless angular momentum \footnote{Indeed, if all binaries have the same $j=j_*$, then the rate is determined by those binaries whose semi-major axis is such that $j_0(\lambda)=j_*$. Assuming, for simplicity, a monochromatic PBH mass spectrum, this determines $\lambda=\lambda_0 = 1.3\cdot 10^{-4} j_*^{-21/16} m^{5/16}.$ Hence, we have
\begin{equation}
\Gamma(t_0;j_*) = {f\rho_{m}\over 4\bar m} \int \delta(t_0 - t[a;j_*]) e^{-X} dX \approx 1.8\cdot 10^4 f^2 {m^{-11/16} \over j_*^{21/16}} Gpc^{-3}yr^{-1} e^{-Y_{min}},
\end{equation}
where we used $t\propto a^4 \propto \lambda^{16/3}$ and $X=\lambda f$ in order to do the integration over $X$ by using the Dirac delta function. This matches Eq. (\ref{onemore}) for $j_*\approx 0.33\ \sigma$.}
 $j_*\sim \sigma$.
 From (\ref{onemore}) we have $\Gamma_M \propto M^{15/16} e^{-Y_{min}}$, and therefore 
 \begin{equation}
 \alpha = {15\over 16} \approx 0.93, \quad (p=0)
 \end{equation}
 which is significantly lower than the values in the range (\ref{alphaA}), corresponding to the cosmological plateau. The difference comes from the fact that for the enhanced spectrum, perturbations in radiation dominate over matter perturbations, and this changes the mass dependence of merger rates. 
 For $P_\Phi \approx 10^{-5}$, and for a monochromatic spectrum, we have
\begin{equation}
\Gamma(t_0) \approx 1.6 \cdot 10^6 f^2 m^{-11/16} e^{-Y_{min}} Gpc^{-3}yr^{-1}.  \quad (p=0)\label{aapprox}
\end{equation}
This turns out to be an excellent approximation to the numerical result which is plotted in Fig. \ref{rateB}. In the mass range $30-100$ solar masses, the LIGO bound on the merger rate requires
$f\sim 5-9 \%$, where we have used $Y_{min}\approx 2$. Hence, the bound on $f$ is significantly relaxed compared to the case without cosmological perturbations.


\begin{figure}
\begin{center}
\includegraphics[scale=0.56]{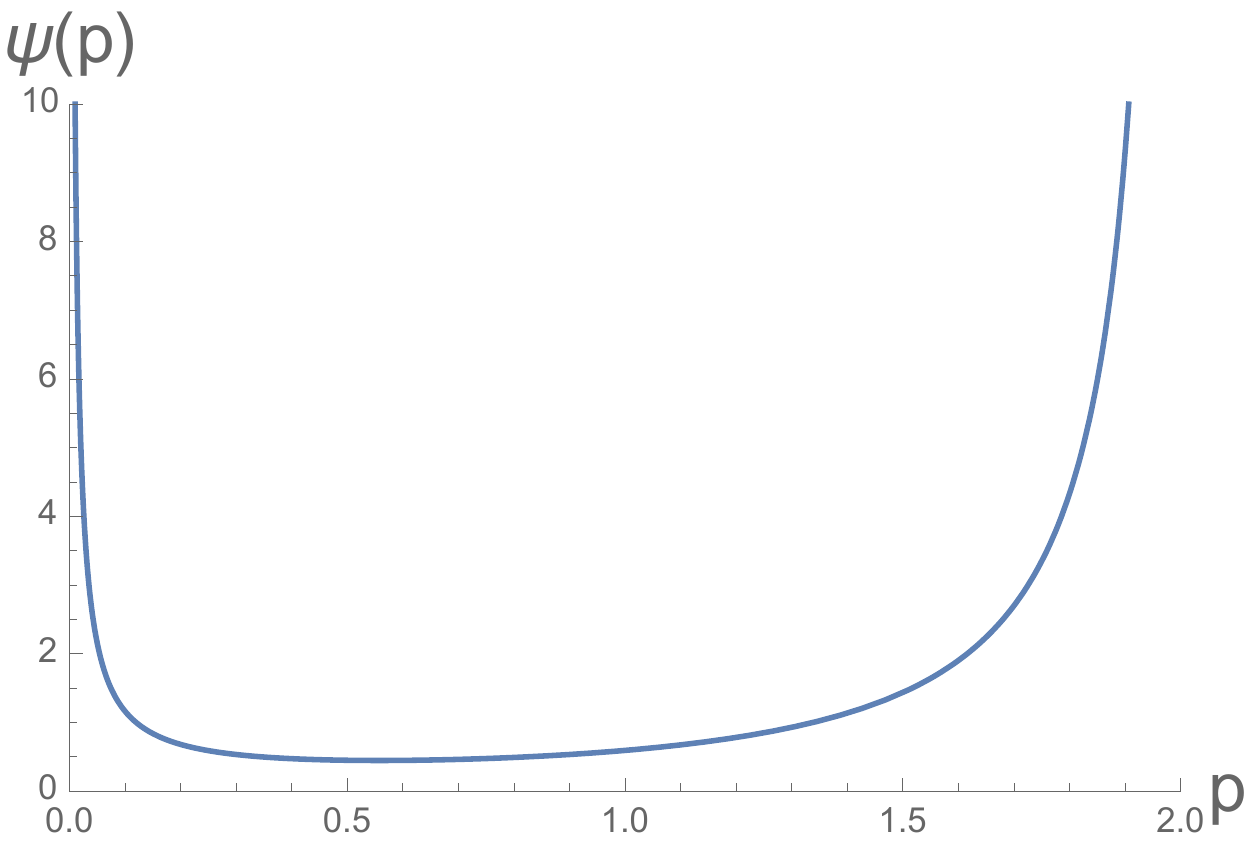}
\caption{The coefficient $\psi(p)$, introduced in Eq. (\ref{psip}), is represented as a function of the spectral index $p$. 
The behaviour near the edges of the interval are due to the logarithmically divergent behaviour of the momentum integral in (\ref{psip}), both for $p=0$ and $p=2$. 
These divergences are regulated by the finite range of wavelenghts contributing to the integral. Near $p=0$, the divergence is infrared and has been taken into consideration in Eq. (\ref{dispplateauB}), where a few orders of magnitude in $k$ contribute to the integral. Consequently, the logarithm is of order $\sim 10$. Similarly, near $p\approx 2$, the behaviour of $\psi(p)$ should be regulated at $\psi(p)\lesssim 10$, since no more than a few orders of magnitude in $k$ will contribute to the corresponding ultraviolet logarighmic divergence, from the scale $k_0$ to the scale 
$k_{max}$ where the power spectrum reaches its maximum value.}
\label{psi}
\end{center}
\end{figure}

For $0<p<2$, the integral in (\ref{totvar}) is dominated by $k \approx k_0$. As we have just seen for the case of $p=0$, when the amplitude of cosmological perturbations is enhanced, it is the perturbations in radiation that dominate, and 
in this case we  know that the contribution of modes with $k\gtrsim k_0$ can also be important. This is given by Eq. (\ref{variancell2}). Defining
\begin{equation}
\psi(p) = \int_{0}^{\infty} {dk \over k} \left({k\over k_0}\right)^p \left[{1-G(kx_0) \over (kx_0)^2}\right] = 3 \cos \left({\pi p\over 2}\right) {\Gamma[p-3] \over p-5},\label{psip}
\end{equation}
and using (\ref{powerB}) in (\ref{variancell2}) we have 
\begin{equation}
\sigma^2_{cp} \approx 2.2\cdot 10^3 \psi(p) \left({k_0\over k_B}\right)^p B_\Phi \left[1+ 3.7 \lambda \left({2\bar m\over M}\right)\right]^2. \label{sigmaBp}
\end{equation}
For the purpose of keeping track of the relative importance of radiation versus matter perturbations, here we have included the effect of matter perturbations up to the time $\eta = \eta_b$, which corresponds to the second term in square brackets in Eq. (\ref{sigmaBp}). Matter perturbations can contribute to the angular momentum for some time after $\eta_b$, when the two PBHs forming the binary start falling towards each other. Here we have not included this contribution, which in principle can have a more complicated mass dependence. However, by order of magnitude, this effect is comparable to the contribution up to the time $\eta_b$. This characterization suffices in order to check whether perturbations in radiation are dominant or not. The numerical coefficient $\psi(p)$ is represented in Fig. \ref{psi}, where we can see that $\psi > 0.44$ for all values of $p$, and that it diverges near both ends of the interval $0<p<2$. Such divergences correspond to the infrared and ultraviolet logarithmic divergences in the integral (\ref{variancell2}), respectively. In practice, these are cut-off by the finite range where the power spectrum has the specified behaviour, as we discussed for the case $p=0$. Here, we shall  simply consider a generic $p \sim 1$, which is not too close to $0$ or $2$, so that $\psi \sim 1$.
In this case, we obtain
\begin{equation}
\sigma^{2\ (p)}_{cp} \approx 0.97\cdot 10^3\  {\psi(p)\over 0.44} \left[1+ 3.7 \lambda  \left({2\bar m \over M}\right)\right]^2 (m\lambda)^{-p/3}\left({10^6 k_{eq} \over 2 k_B}\right)^p B_\Phi. \quad (p \sim 1) \label{sigmaBp2}
\end{equation}

In order to estimate the merger rates, let us first start by neglecting matter perturbations in (\ref{sigmaBp2}). Using (\ref{j0}), we have 
\begin{equation}
{j_0^2\over \sigma^2}  \approx \left({\lambda\over \lambda_B^{(p)}}\right)^{7p-32\over 21}, \label{zoo}
\end{equation}
with
\begin{equation}
\lambda_B^{(p)} = \left[{2.7 \cdot 10^{-9} \over B_\Phi}\left({0.44\over \psi(p)}\right) \left({M\over 2 \bar m}\right)^2 \left({ 4 m_1 m_2 \over M^2}\right)^{2/7} \right]^{21\over 32-7p} m^{10+7p\over 32-7p},
\end{equation}
where we have set $k_B= 5\cdot 10^5 k_{eq}$. For $B_\Phi \gtrsim 10^6$ and $m\lesssim 100$, we find that $\lambda_{B}^{(p)} \ll 1$. This is consistent with the assumption that perturbations in 
radiation dominate over perturbations in matter, since the peak of the differential rate $W(\lambda)$, given by (\ref{wlambda}), occurs at $\lambda \sim \lambda_B^{(p)}\ll1$. 

\begin{figure}
\begin{center}
\includegraphics[scale=0.7]{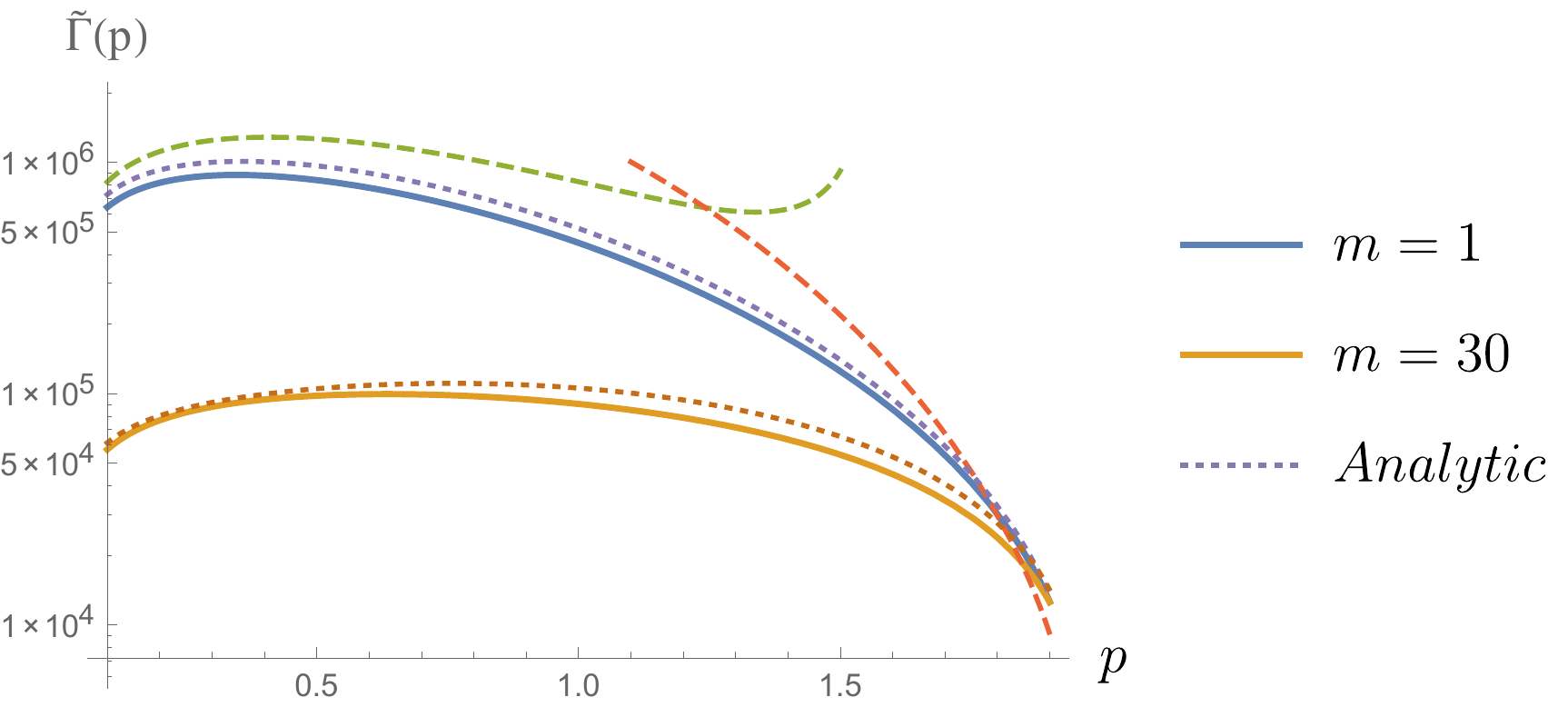}
\caption{The merger rate $\tilde \Gamma(p)$ with the $f$ and $Y_{min}$ dependence factored out, as in Eq. (\ref{gbp}), for $B_\Phi=10^{-5}$. The dashed lines represent the approximations (\ref{dim}) and (\ref{dim2}), where the torque from non-relativistic matter perturbations is neglected. Here, we use the value $m=1$. The thick lines correspond to the numerical evaluation of the rate, for $m=1$ and $m=30$, including the torque of matter perturbations up to the time $\eta=\eta_b$ when the binary decouples from the Hubble flow. The dotted lines correspond to the analytic estimate given in Eq. (\ref{gammap}), which is a very accurate approximation. All curves are for a monochromatic mass distribution.}
\label{caseBp}
\end{center}
\end{figure}

For $0\lesssim p\lesssim 1$, the integral in (\ref{intrate}) quickly converges for $\lambda \gg \lambda_B^{(p)}$ and the upper limit of integration becomes irrelevant. In this case we have 
\begin{equation}
\Gamma_M(t_0) \approx {\rho_m \over M_\odot t_0} {f^2 \over 14 m} \left({21\over 32-7p}\right)\Gamma\left({11-7p\over 32-7p}\right) \lambda_B^{(p)} e^{-Y_{min}}.\quad \left({0\lesssim p \lesssim 1}\right). \label{dim}
\end{equation}
As $p$ approaches the value $11/7$, the  
ratio given in (\ref{zoo}) approaches the behaviour $(j_0^2 / \sigma^2)\propto \lambda^{-1}$, and the integral (\ref{intrate}) becomes logarithmically divergent at large $\lambda$. This is regulated by the finite range of $\lambda$. Indeed, for $1\lesssim p \lesssim 2$, the integral (\ref{intrate}) is dominated by the interval $\lambda_{B}^{(p)} \lesssim \lambda< \lambda_{m}$. Here, $\lambda_{m}\sim 0.3$ is the value for which matter perturbations start becoming important. 
At that point, the behaviour of the integrand switches to $(j_0^2 / \sigma^2)\propto \lambda^{-3+{7 p-11\over 21}}$, which rapidly converges. Hence, in that case we can approximate
\begin{equation}
\Gamma_M(t_0) \approx {\rho_m \over M_\odot t_0} {f^2 \over 14 m} \ln\left[\lambda_m/\lambda_B^{(p)}\right] \lambda_B^{(p)} e^{-Y_{min}},\quad \left(1\lesssim p \lesssim 2\right).\label{dim2}
\end{equation}
Let us now consider the "universality" coefficient $\alpha$, defined in Eq. (\ref{alphadef}). From (\ref{dim}) or (\ref{dim2}), we immediately obtain
\begin{equation}
\alpha\approx -M^2 \partial^2_{m_1,m_2}\ln[\lambda_B^{(p)}] \approx {30\over 32-7p},\quad  (0\lesssim p \lesssim 2) \label{alphaBp}
\end{equation}
where in the case (\ref{dim2}) we have ignored the subleading logarithmic dependence on $M$ (we will comment on this subleading correction below).

\begin{figure}
\begin{center}
\includegraphics[scale=0.7]{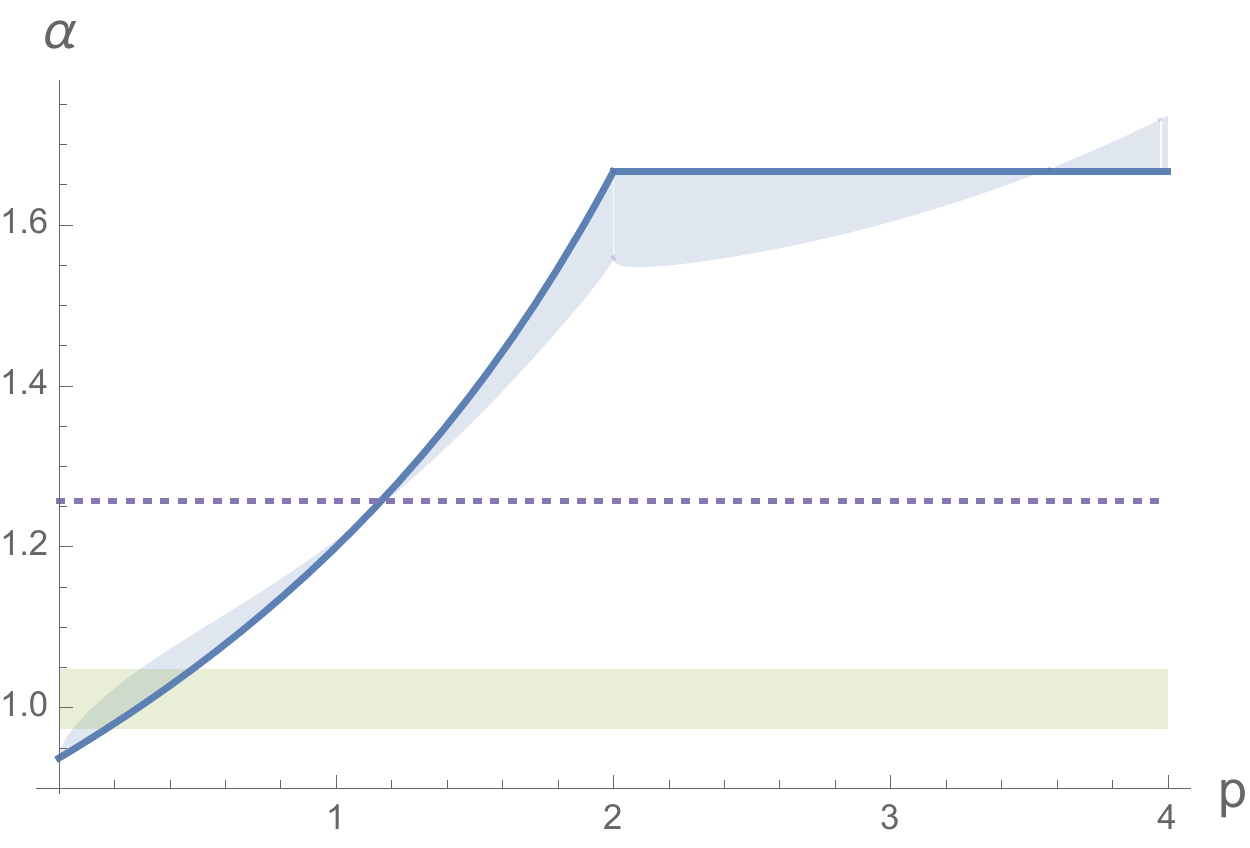}
\caption{The ``universality" coefficient $\alpha$, defined in (\ref{alphadef}), is plotted as a thick line for an enhanced power spectrum with spectral index $p$ and amplitude $B_\Phi\gtrsim 10^{-6}$ at scales comparable to the co-moving size of the binaries. This coefficient is given by Eq. (\ref{alphaBp}) in the range $0\lesssim p \lesssim2$. For comparison, the case where cosmological perturbations are neglected and the torque of binaries is entirely due to neighboring PBHs corresponds to the horizontal shaded interval, given in Eq. (\ref{unitorque}), or to the horizontal dotted line when binaries for which the nearest PBH has $Y<Y_{min}\sim 2$ are excluded from the count. The latter are likely to be severely  affected by infalls, as discussed around Eq. (\ref{alphamany}). The shade around the thick line corresponds to the correction $\Delta \alpha$ given in Eq. (\ref{deltaalpha}), where we have taken $\bar m= 20$ and $m_1\sim m_2 \sim 20$. For $2\lesssim p \lesssim 4$, and  $\lambda_C \ll 1$ we have $\alpha \approx 5/3$ [see the discussion around Eq. (\ref{lowkmax})]. The shade around the horizontal line $\alpha \approx 5/3$ corresponds to $\Delta \alpha$ taken from the analytic estimate (\ref{tildegammaC}), with $C_\Phi \approx 10^{-3}$ and $k_C = 10^7 k_{eq}$.}
\label{alpha}
\end{center}
\end{figure}

Factoring out the $f^2$ and $Y_{min}$ dependence in $\Gamma_M$, as
\begin{equation}
\Gamma_M \approx \tilde\Gamma(p) f^2 e^{-Y_{min}} Gpc^{-3} Yr^{-1}, \label{gbp}
\end{equation}  
we have represented the approximations (\ref{dim}) and (\ref{dim2}) in Fig. \ref{caseBp} as dashed lines (for the case $m=1$). For comparison, we numerically calculate the rate which is obtained when we include, in addition to the effect of perturbations in radiation, the torque exerted by non-relativistic matter perturbations up to the time $\eta\approx \eta_b$ when the binary decouples from the Hubble flow. The latter torque corresponds to the second term in square brackets in Eq. (\ref{sigmaBp}). The numerical result for $m=1$ and $m=30$ is depicted in thick lines in Fig. \ref{caseBp}.  Note that the approximations (\ref{dim}) and (\ref{dim2}) can be imprecise, by up to a factor of 2 or so in the region $p\sim 1$, due to the fact that we have neglected the effect of matter perturbations, which can affect the merger rates by a sizable fraction even if they are subdominant. 

\begin{figure}
\begin{center}
\includegraphics[scale=0.56]{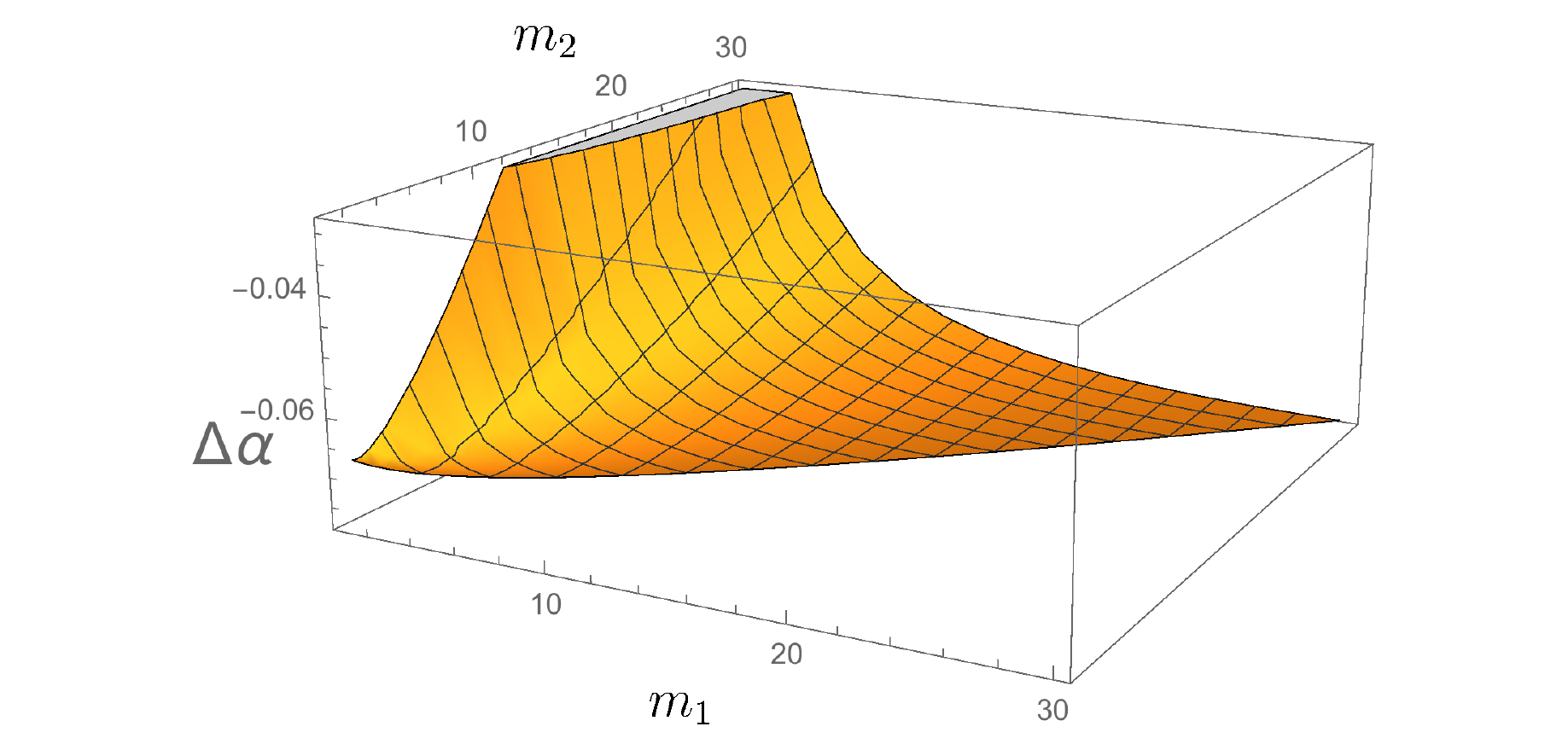}
\caption{The  correction $\Delta\alpha$ in Eq. (\ref{deltaalpha}), for $p=1.8$, and $\bar m = 10 M_\odot$, as a function of $m_1$ and $m_2>m_1$ (expressed in solar masses). 
The correction is small unless the mass ratio is hierarchical ($m_1\ll m_2$), corresponding to the region near the left boundary of the plot.}
\label{twomasses}
\end{center}
\end{figure}

A much better analytic approximation to the numerical result can be obtained by keeping the effect of matter perturbations in $\sigma^2_{cp}$. Using $\lambda_B^{(p)} \ll 1$, we may approximate
\begin{equation}
\int_0^{2M\over \bar m} W d\lambda \approx  2 \int_{\lambda_{B}^{(p)}}^\infty  \left({\lambda\over \lambda_B^{(p)}}\right)^{7p-32\over 21} \left[1+3.7 \lambda \left({2\bar m\over M}\right) \right]^{-2} d\lambda.
\end{equation}
Here, we have neglected the integrand for $\lambda \lesssim \lambda_B^{(p)}$, since this is a small interval where $W$ is suppressed relative to its peak value at $\lambda \sim  \lambda_B^{(p)}$, and we have also neglected the exponential factor in $W$ for $\lambda \gtrsim \lambda_B^{(p)}$, since the exponent is small in this range. The integral in the right hand side can be calculated in terms of the incomplete Euler's $\beta$ function. Expanding this function for small $\lambda_B^{(p)}$ we find
\begin{equation}
\tilde\Gamma(p) \approx {\rho_m \over M_\odot t_0} {\lambda_B^{(p)} \over 14 m} \left({21\over 11-7p} + \left[3.7\left({2\bar m\over M}\right)\lambda_B^{(p)} \right]^{11-7p \over 21}{\pi {32-7p\over 21}\over \sin\left(\pi {32-7p\over 21}\right)}\right). \label{gammap}
\end{equation}
The analytic approximation (\ref{gbp}) with (\ref{gammap}) is plotted in Fig \ref{caseBp} in dotted lines for $m=1$ and $m=30$. We find that it reproduces the full numerical result plotted in thick lines, with very good accuracy in the full range of $p$.
The term in round brackets contains additional dependence on $m_1$ and $m_2$ which is not present in (\ref{dim}), and this will contribute a correction $\Delta\alpha$ to the expression (\ref{alphaBp}), given by
\begin{equation}
\Delta\alpha = - M^2 \partial^2_{m_1,m_2} \ln\left|\left[3.7\left({2\bar m\over M}\right)\lambda_B^{(p)} \right]^{11-7p \over 21}{ (32-7p)\over 21\ {\rm sinc\left(\pi {11-7p\over 21}\right)}}-1\right|. \label{deltaalpha}
\end{equation}
This correction turns out to be rather small, and typically $\Delta \alpha \lesssim 0.05- 0.1$, unless the mass ratio is hierarchical. The reason can be understood as follows. The mass dependence comes from the term in square brackets in (\ref{deltaalpha}), 
\begin{equation}
\left[\left({2\bar m\over M}\right)\lambda_B^{(p)} \right]^{11-7p \over 21} \propto M^{\beta} (m_1m_2)^\gamma. \label{betagamma}
\end{equation}
The exponents $\beta$ and $\gamma$ 
happen to be very small $\beta, \gamma \lesssim 0.1$ in the range of interest. Then, unless one of the two masses is hierarchically smaller than the total mass $M$, we have $\Delta \alpha \propto \beta + O(\beta^2 ,\beta\gamma,\gamma^2) \sim \beta$. The correction $\Delta \alpha$ is plotted in Fig. \ref{twomasses} for $p=1.8$, $\bar m = 10M_\odot$, and a range of values of $m_1$ and $m_2$. Note that $|\Delta \alpha| \lesssim 0.08$, unless we consider the region where $m_1\ll m_2$. 
We have checked that for the case of similar masses $m\lesssim 30$, we have $|\Delta \alpha|\lesssim 0.08$ in the full range $0.2\lesssim p \lesssim 1.8$


We conclude that, in the range $0\lesssim p \lesssim 2$ the spectral index $p$ determines the parameter $\alpha$, which may range from $15/16\approx .93$ up to $ 5/3 \approx 1.66$. This is represented in Fig. \ref{alpha}.
A measurement of $\alpha$ may therefore provide valuable information on the primordial perturbation power spectrum at intermediate scales. Note that the value of $\alpha$ is independent of the amplitude of perturbations, as long as this amplitude is sufficiently large \footnote{At low amplitudes, $B\lesssim 10^{-7}$ the torque from matter perturbations becomes as important as that from radiation perturbations. In general, in this case, the dependence of $\Gamma_M$ on $M$ does not necessarily factor out as a power of $M$. Still, if we make the assumption that the torque from matter perturbations is dominated by the contribution from times up to $\eta_b$, then  using (\ref{sigmaBp}) and neglecting the first term in square brackets (which is due to radiation perturbations), we find $\alpha = 72/(74-7p)$, which is larger than $36/37\approx 0.97$ and smaller than $6/5 = 1.2$. This is a narrower range than allowed by perturbations in radiation.}
$B_\Phi \gtrsim 10^{-6}$.

It should also be pointed out that the bound on $f$ from the observed merger rates can be considerably relaxed by the enhanced power spectrum at intermediate scales. For instance, if we take $m=30$ and $B_\Phi = 10^{-5}$, then from 
Fig. \ref{caseBp} we have $\tilde \Gamma(p)\lesssim 10^{5}$. Consequently, for a monochromatic PBH mass function, the bound
\begin{equation}
\Gamma(t_0) \approx \tilde\Gamma(p) f^2 e^{-Y_{min}}  Gpc^{-1} Yr^{-1} \lesssim 10^2 Gpc^{-1} Yr^{-1}, 
\end{equation}
with $Y_{min}\approx 2$ is satisfied for $f\lesssim 0.1$. \footnote{For smaller values of the power spectrum, in the range $10^{-7}<B_\Phi<10^{-5}$, the merger rate scales approximately in proportion to $\lambda_B^{(p)}\propto B_\Phi^{-21/(32-7p)}$. This translates into the bound $f \lesssim 0.1 (B_\phi/10^{-5})^{21/(64-14p)}$.}

\subsection{Case C: Steep spectrum at very small scales.} \label{bump}

Consider a power spectrum with spectral index $p\gtrsim 2$ up to some high scale $k_C$. 
In this case, the contribution of perturbations to the angular momentum of binaries will be dominated by $k\sim k_C$, rather than $k\sim k_0$,
and it is convenient to parametrize by using $k_C$ as the pivot scale:
\begin{equation}
P_\Phi(k) = C_\Phi \left({k\over k_C}\right)^p.  \quad (p>2, \quad k\lesssim k_C) \label{powerC}
\end{equation}
From the constraints in Fig. (\ref{power}), and assuming $k_C\geq 10^7 k_{eq}$, we require
\begin{equation}
C_\Phi\equiv P_\Phi(k_C)\lesssim 10^{-2}-10^{-3}. \label{cbound}
\end{equation}
The case with $p<2$ was discussed in the previous subsection, with the correspondence 
\begin{equation}
C_\Phi = B_\Phi \left({k_C\over k_B}\right)^p,
\end{equation}
between the prefactors in (\ref{powerB}) and (\ref{powerC}).
Here we concentrate in the case $p>2$. The specific form of $P_\Phi(k)$ for $k\gtrsim k_C$ will not be important, as long as it grows slower than $k^2$, or that it decays for $k>k_C$.

From (\ref{variancell2}), the contribution of radiation perturbations to the variance of $j$ is dominated by wavelengths shorter than the binary size $x_0$, and is given by
\begin{equation}
\sigma^2_{cp (rad)} \approx 2.2\cdot 10^3 \int_{k_0}^{k_C} {dk\over k}  P_\Phi(k) {1\over (k^2 x_0^2)} \approx {2.2\over p-2}\left({C_\Phi\over 10^{-3}}\right) \left({k_0 \over k_C}\right)^2.
\end{equation}
In the last step we have neglected the contribution from the lower limit of integration, assuming that $p$ is not too close to $2$. As we discussed in the previous subsection, we can easily incorporate 
the torque of matter perturbations up to the time 
$\eta=\eta_b$, which gives a combined total of
\begin{equation}
\sigma^2_{cp} \approx {2.2 \over p-2} \left({C_\Phi\over 10^{-3}}\right)\left({k_B \over k_C}\right)^2  (m\lambda)^{-2/3} \left[1+3.7\lambda\left({2\bar m\over M}\right)\right]^2. \label{sigmacpC}
\end{equation}
Here we have used $k_0/k_B = (m\lambda)^{-1/3}$, with $k_B= 5\cdot 10^5 k_{eq}$.

Given that $C_\phi$ is bounded above by (\ref{cbound}), the factor $(k_B/ k_C)^2$ will considerably suppress the effect of cosmological perturbations unless $k_C$ is not too far from the intermediate scale $k_B$.
Eq. (\ref{sigmacpC}) can be compared with de contribution of neighboring PBH to the torque, given in (\ref{varnb})
\begin{equation}
\sigma^2_{nb} \approx {K \over 4Y_{min}} f^2 \lambda^2\left({2\bar m\over M}\right)^2 \sim 10^{-1} \lambda^2 f^2,  \label{varnb2}
\end{equation}
where in the last step we use the fiducial values $K\sim 1$, $M\sim 2\bar m$ and $Y_{min}\approx 2$.  
For $p$ not too close to 2,  and $m\sim 30$, the effect of cosmological perturbations can only be dominant for a narrow range of $k_C$:\footnote{In the case of cosmological perturbations, the merger rate for $p>2$ is dominated by the values of $\lambda$ where the behaviour of the variance (\ref{sigmacpC}) changes from $\sigma^2_{cp} \propto \lambda^{-2/3}$ to $\sigma^2_{cp} \propto \lambda^{4/3}$. This happens at $\lambda \sim 1/3$, and we use this in the relation $\sigma_{cp}\gtrsim \sigma_{nb}$ in order to estimate the upper limit of the range (\ref{restrictive}).} 
\begin{equation}
10^7 k_{eq} \leq k_C \lesssim 2\cdot 10^7 f^{-1} (p-2)^{-1/2} m^{-1/3} k_{eq}.\label{restrictive}
\end{equation}
Hence, the existence of a window for $k_C$ where $\sigma_{nb} \ll \sigma_{cp}$ requires $f\ll 1$.

\begin{figure}
\begin{center}
\includegraphics[scale=0.7]{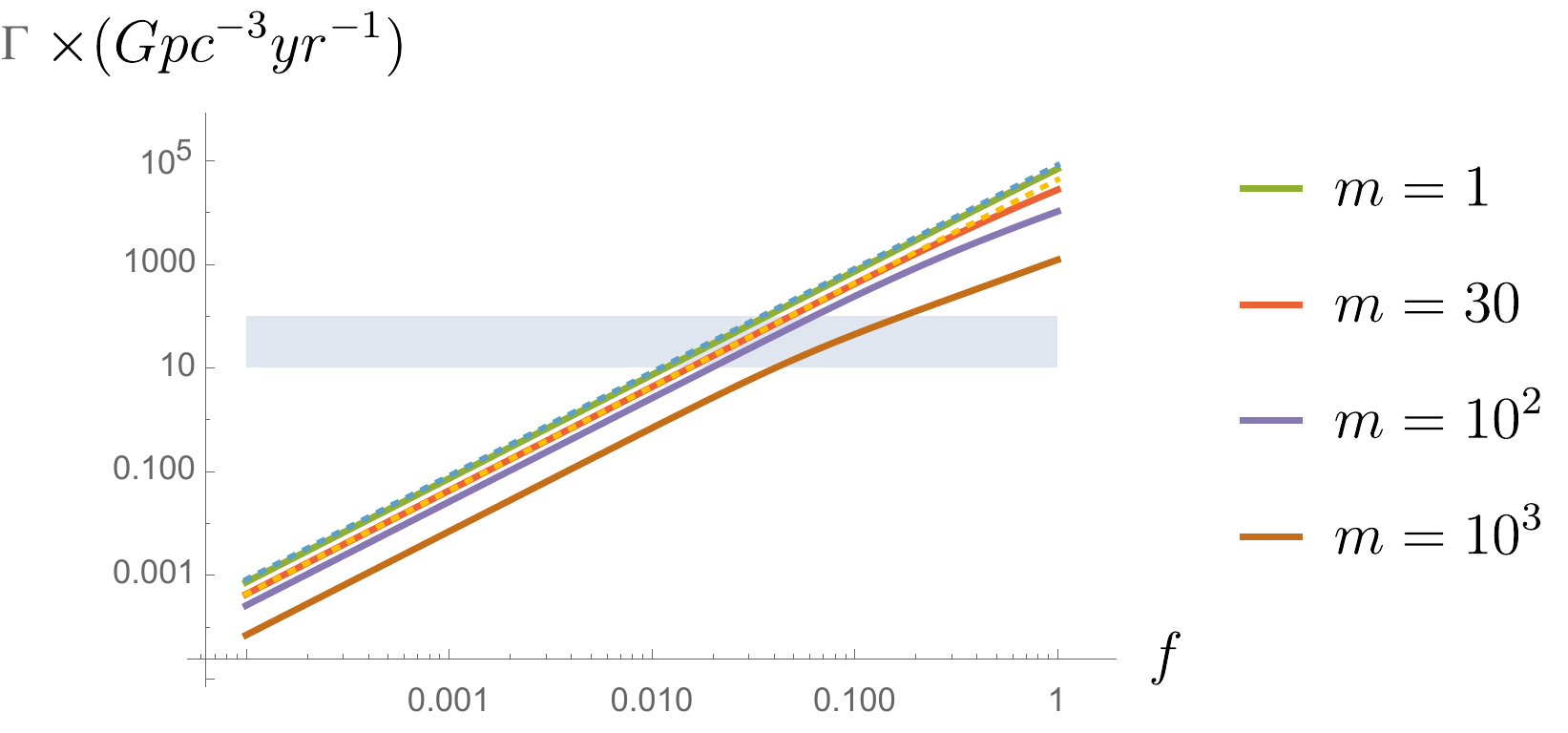}
\caption{The merger rates for Case C, with power spectrum amplitude $C_\Phi = 10^{-3}$ at $k_C= 10^7 k_{eq}$, and with $p=4$. The thick lines correspond to the numerical evaluation of the rates for different values of the mass, while the 
dashed lines correspond to the analytic estimate (\ref{tildegammaC}), which is  a very good approximation for $m\lesssim 30$. Note that the rate has only a very mild dependence on the mass, at least up to $m\lesssim 10^2$. Despite a large amplitude of primordial perturbations, the effect on the merger rate is not as significant as in Case B, since now the perturbations are on scales smaller than the binary size.}
\label{caseC2}
\end{center}
\end{figure}

Following similar steps as in Subsection \ref{intermediate}, we find that the expected merger rate of binaries will be given by (\ref{gbp}), with
\begin{equation}
\tilde\Gamma(p) \approx {\rho_m \over M_\odot t_0} {\lambda_C \over 14 m} \left( \left[3.7\left({2\bar m\over M}\right)\lambda_C \right]^{-1/7}{6\pi/7\over\sin\left(6\pi/7 \right)}-7\right).  \quad (\lambda_C \ll 1) \label{tildegammaC}
\end{equation}
Here,
\begin{equation}
\lambda_C \equiv \left[1.2 (p-2) 10^{-6}\left({k_C\over k_B}\right)^2 \left({10^{-3}\over C_\Phi}\right)\right]^{7/6}  \left({M\over 2 \bar m}\right)^{7/3} \left({ 4 m_1 m_2 \over M^2}\right)^{1/3}m^{4/3}.
\end{equation}
This leads to a value of the universality coefficient which is given by (\ref{alphaBp}) with $p=2$. That is
\begin{equation}
\alpha \approx {5\over 3}.  \quad (\lambda_C \ll 1) \label{lowkmax}
\end{equation}
Eq. (\ref{lowkmax}) corresponds to the thick horizontal line in the range $p>2$ in Fig. \ref{alpha}. Note, however, that the approximation (\ref{tildegammaC}) requires $\lambda_C \ll1$ (since it is obtained by Taylor expanding an incomplete Euler beta function in small $\lambda_C$). The merger rates for $k_C = 10^7 k_{eq}$ are plotted in Fig. \ref{caseC2}, where it can be seen that the approximation (\ref{tildegammaC}) plotted as dashed lines, agrees very well with the numerical result plotted in thick lines, for $m\lesssim 30$. For higher values of $k_C$ or higher values of $m$, we are outside the regime $\lambda_C \ll 1$, and the approximation is not valid. Nonetheless, the behaviour of the merger rates 
from the numerical result is similar for all masses up to $m\lesssim 100$. In particular, we find that $f$ should be less than a few percent in the whole range $m = 1-100$. This is in contrast with the results for Case B ($p<2$), where values of $f$ as large as $10\%$ can be consistent with the observed merger rates. In spite of the high amplitude of the power spectrum, the effect in Case C is not as significant as in case B, since perturbations are now on scales smaller than the binary size.

\subsection{Effect of a peak at the PBH scale}

The mechanism of PBH formation from adiabatic perturbations relies on a prominent enhancement in $P_\phi$, with a maximum at some co-moving scale $k_{PBH}$ with amplitude 
\begin{equation}
P_\Phi(k_{PBH}) \sim 10^{-3}-10^{-2}. \label{powerpbh}
\end{equation}
This is enough to produce PBHs in significant abundance, in regions where high peaks in the Gaussian random field of cosmological perturbations reach non-linear values above a certain threshold. As discussed in Refs. \cite{YHGK,GM}, PBH formation does not occur when $k_{PBH}$ crosses the horizon, but slightly later, when the scale 
\begin{equation}
r_m \sim 3\ k_{PBH}^{-1} \label{rm}
\end{equation}
crosses the horizon\footnote{The scale $r_m$ corresponds to the maximum of the so-called compaction function ${\cal C}(r)$, which characterizes the averaged overdensity as a function of the distance to its center. 
The approximate factor of 3 in (\ref{rm}) depends somewhat on the shape of the overdensity profile, which for amplitudes well above the standard deviation is in turn determined by the form of $P_\Phi(k)$. For instance, for a monochromatic power spectrum,  
$P_\Phi(k) \propto \delta(k-k_{PBH})$, we have $r_m \approx 2.7\  k_{PBH}^{-1}$. On the other hand, the strong overdensity causes a non-linear distortion of the spatial metric which affects the relation between co-moving and physical distance. Generically,
the combination of these non-linear effects gives a mass of the black hole which is $\sim 10$ times bigger than the mass within the unperturbed horizon at the time when $k_{PBH}$ crosses it. Hence the factor of 3 in (\ref{PBHm}).}. The mass of the black hole is typically equal to the mass within the horizon at that time, so that $2GM=H_{r_m}^{-1} $, and we have  
\begin{equation}
k_{PBH} \approx 3\cdot 10^9 m^{-1/2} k_{eq}.\label{PBHm}
\end{equation}
Comparing with Eq. (\ref{k0}) this corresponds to lengthscales which are shorter than the co-moving size of the binary, by three orders of magnitude or so, corresponding to the present co-moving wave-number of order 
\begin{equation}
\bar k_{PBH} \sim 3\cdot 10^{7} m^{-1/2} Mpc^{-1}.
\end{equation}
On a logarithmic scale, this is not too far from the intermediate scales discussed in Subsection \ref{intermediate}, and hence it is natural to ask about the consequences of this peak on the merger rates, in the light of our earlier results. 


For illustration, in Fig. \ref{power} we include an example extracted from \cite{BCP}, where the enhancement of the power spectrum is obtained from a period where the inflaton goes from slow roll to ultra-slow roll and back to slow roll, through a discrete set of values of the second slow roll parameter $\eta$. This model may be somewhat artificial, but it will be useful for illustrative purposes since it includes several features which may be present in more general cases. The curve includes a steep growth with $P_\Phi \propto k^p$ from the cosmological plateau up to a very high value $P_\Phi\sim 10^{-3}$, at $k\sim k_C$, as in Case C. This is followed by a less steep part up to some maximum value $k_{PBH}$, corresponding to the scale of PBH formation. Finally, the spectrum decays for $k>k_{PBH}$.

It has been argued that $p=4$ corresponds to the theoretical upper bound for the spectral index for a wide class of inflationary one field models \cite{BCP}. Values in the range 
$3\lesssim p \lesssim 4$ are easily obtained in the case where the enhancement is due to a short period of ultra-slow roll and constant roll inflation. The upper bound can be saturated in relatively simple models where the constant roll phase has a sufficiently negative value of the slow roll parameter $\eta$ (see e.g. \cite{AGM}).
For $p\gtrsim 2$, the regime where cosmological perturbations dominate over the effect of neighboring PBHs is given by Eq. (\ref{restrictive}), which requires
\begin{equation}
f\lesssim 
6.7\cdot 10^{-3} {m^{1/6}\over \sqrt{p-2} }\left({k_{PBH}\over k_C}\right). \quad (p \gtrsim 2) \label{condisteep}
\end{equation}
Here we have used (\ref{PBHm}). 
For a sizable value of the fraction of dark matter in the form of PBHs, say $f\gtrsim 10^{-2}$, cosmological perturbations will only play a role if the peak of the power spectrum is rather {\em broad}, with $k_{PBH} \gtrsim 10\ k_C$. Conversely, for a {\em narrow} peak with $k_{PBH} \sim k_C$ and sizable $f\gtrsim 10^{-2}$, the eccentricity of binaries is determined by the tidal forces from neighboring PBHs. Broad peaks in the power spectrum are not necessarily generic, but as shown by the example in Fig. \ref{power}, they can in principle be obtained in phenomenological models. 

\begin{figure}
\includegraphics[scale=0.5]{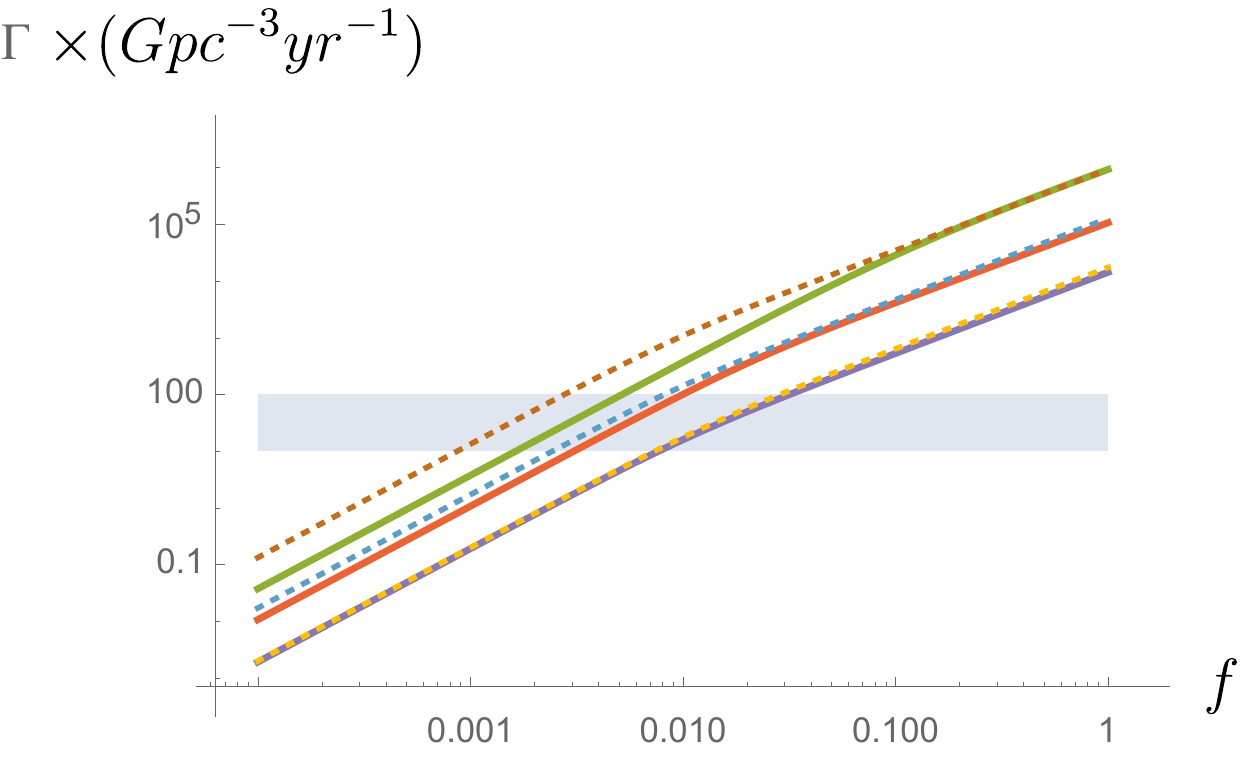}
\includegraphics[scale=0.5]{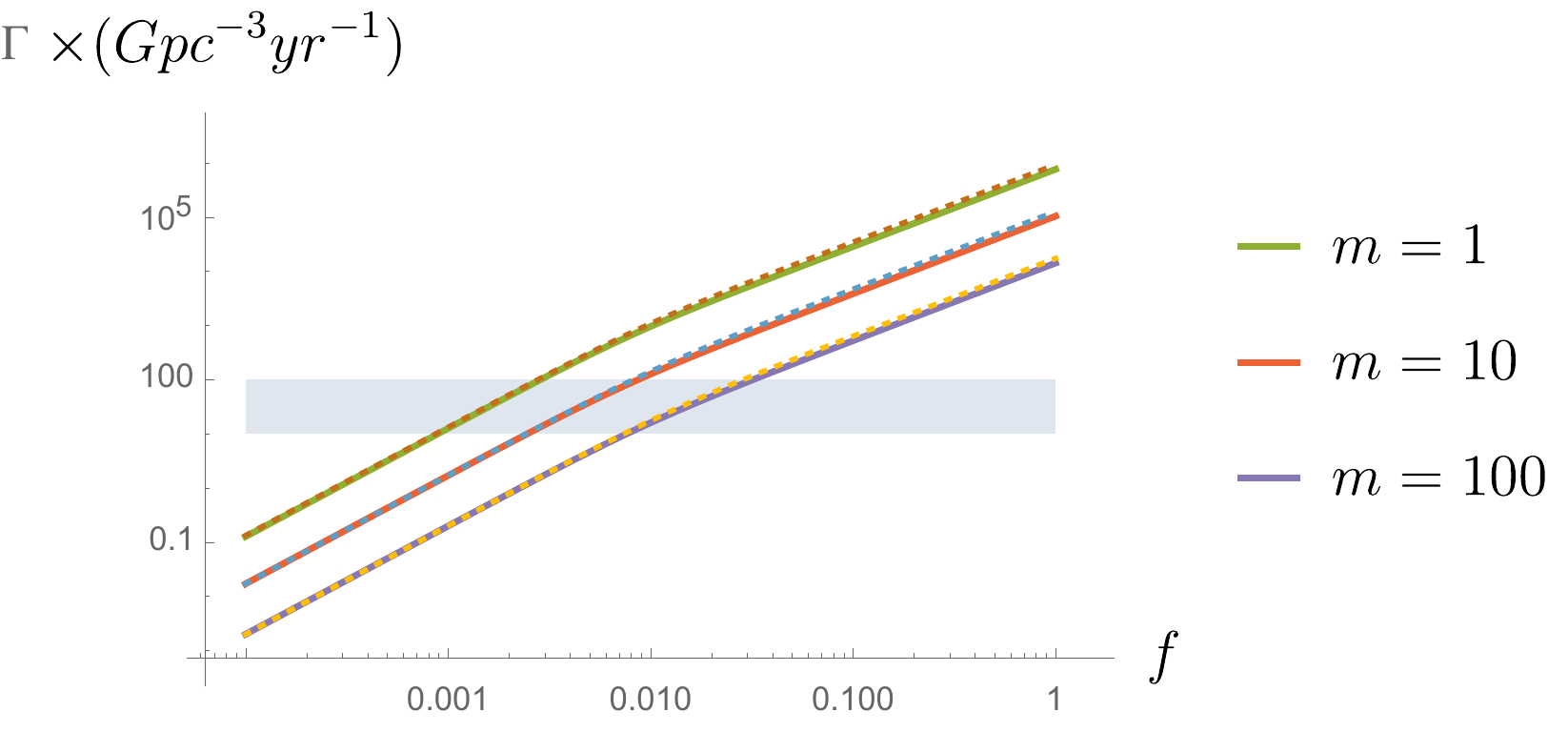}
\caption{The thick lines in the left pannel represent the merger rates for a steep power spectrum of cosmological perturbations with $p=4$, and a broad peak spanning from $k_c\approx 10^8 k_{eq}$ to $k_{PBH} \approx 10^9 k_{eq}$, with amplitude $C_\Phi \approx 10^{-3}$ and different values of the mass. Since the peak only affects very small scales and has no power at intermediate scales, we have added a cosmological plateau with normalization $P_\phi\approx 10^{-9}$, representing a minimal contribution we may expect for $k< 10^7 k_{eq}$. For reference, the dashed lines represent the case where only the flat plateau with cosmological normalization is taken into consideration, as in Case A. The effect of the steep bump can be significant for low mass PBHs $m\sim 1$, but not for higher masses.
In the right pannel, we consider the case with $k_C=k_{PBH} = 10^9 k_{eq}$, corresponding to a sharp peak at the PBH scale. The effect of the steep bump is completely negligible in this case.
}
\label{caseC3}
\end{figure}

The merger rate of binaries is plotted in the left pannel of Fig. \ref{caseC3} for different values of the mass, for the case of a broad peak which raises steeply with $p=4$ up to the scale $k_C\approx 10^8 k_{eq}$, and then proceeds with moderate slope to the maximum at $k_{PBH}=10 k_C \approx 10^9 k_{eq}$. In this case the analytic approximation (\ref{tildegammaC}) does not apply, since $\lambda_C \gtrsim 1$ in the relevant range of $f$. The effect of cosmological perturbations is only significant for low masses $m\lesssim 1$.
For the case of a narrow peak with $p=4$, which raises from low values with a steep spectral index to all the way to the PBH scale $k_{PBH}=k_C \approx 10^9 k_{eq}$, the merger rate is plotted in the right pannel of Fig. \ref{caseC3}, where we see that the effect of the narrow peak is completely insignificant.

For less steep spectral index $0<p\lesssim 2$, the growing part of the spectrum corresponds to Case B. The condition that cosmological perturbations provide the dominant source of torque on binaries can be derived along similar lines, and is given by
\begin{equation}
f\lesssim 
2^{1-p}\ 10^{1-{3p\over 2}}  m^{p/12} \left({k_{PBH}\over k_C}\right)^{p/2}. \quad (0\lesssim p \lesssim 2) \label{condiflat}
\end{equation}
This is much less restrictive on $f$ than (\ref{condisteep}). Indeed, for $p\lesssim1$, the condition does not significantly restrict the fraction $f$, even for the case where $k_C \approx k_{PBH}$.
\footnote{Note, however, that in order to interpolate between the cosmological value $P_\Phi \sim 10^{-9}$ and the peak value $P_\Phi \sim 10^{-3}$ in the span of wavelengths which goes from $10 Mpc^{-1}$ to $10^7 Mpc^{-1}$, we need
 $p\gtrsim 1$. Incidentally, for $p=1$ the value of the ``universality" parameter $\alpha$ which is obtained when cosmological perturbations provide the dominant torque (see Fig. \ref{alpha}) is quite similar to the value $1.26$ which corresponds to a dominant torque from neighboring PBHs (at a distance larger than the infall radius $Y_{min}$), so for this particular value of $p$ it seems harder to distinguish one mechanism from the other.} The effect of such a feature in the power spectrum is described in Subsection \ref{intermediate}.

\section{Summary and conclusions}
\label{conclusions}

We have studied the effect of cosmological perturbations on the merger rate of PBHs in the stellar mass range. The effect can be quite significant depending on the amplitude of the power spectrum at different scales. 

For a scale invariant spectrum with amplitude $P_\Phi\approx 10^{-9}$ (Case A), matter perturbations have a dominant effect on the eccentricity of binaries  for $f\lesssim 10^{-2}$, while for larger $f$ the distribution of $j$ is actually dominated by tidal torques from PBHs in the vicinity. In this case, the merger rates would be greater than the current LIGO/Virgo bounds unless the fraction of dark matter in the form of PBH is rather small, of the order $f\lesssim 10^{-2}$ for PBHs with $m\sim 30$. 

On the other hand, $P_\Phi$ could be much larger at scales $10-10^5 Mpc^{-1}$ (Case B). For instance, we find that a nearly flat spectrum with amplitude $P_\Phi \gtrsim 10^{-7}$ within such scales leads to a dimensionless angular momentum with mean squared value 
\begin{equation}
\langle j_{cp} ^2 \rangle\sim 10^3 P_\Phi,  \label{flatP}
\end{equation} 
which is mostly due to perturbations in the radiation fluid.
For a nearly flat $P_\Phi$, the variance of $j$ is almost independent of the mass of the PBH and the size of the binaries. There is only a mild subleading logarithmic dependence on such parameters, which accounts for the range of scales contributing to the effect, from the co-moving size of the binary to the co-moving size of the horizon at the time when the binary forms.  The situation is different when we have a tilted spectrum, of the form
$P_\Phi \propto k^p$, ($p\lesssim 2$), in which case the variance of $j$ depends on binary size and masses. In this case the effect is determined by the amplitude of $P_\Phi$ near the co-moving scale of the binaries $k\sim k_0$,
which is in the intermediate range $10^{3}-10^{5} Mpc^{-1}$. 
The bound from spectral distorsions in the CMB caused by dissipation of acoustic modes requires $P_\Phi \lesssim 10^{-5}$ at such scales \cite{Chluba}. In the situation where this bound is saturated, the torque can be large enough to significantly suppress the merger rate to a level consistent with LIGO/Virgo observations even for $f\sim 10\%$.

An even stronger enhancement of the power spectrum may occur at scales beyond $10^{5} Mpc^{-1}$. These are smaller than the co-moving size of the binaries, but they can contribute to the peculiar velocities of the PBHs and hence to the initial orbital angular momentum. For a steep power spectrum (Case C), with spectral index $p\gtrsim 2$ up to some short wavelength $k_C\gtrsim 10^7 k_{eq}$, the contribution to $\langle j^2\rangle$ is dominated by the shortest scales $k\sim k_C$, rather than $k\sim k_0$. The observational upper bound $P_\Phi(k_C) \lesssim 10^{-3}-10^{-2}$ is comparable to the amplitude which is necessary to produce PBH in significant abundances by the collapse of adiabatic perturbations.  For generality, we may assume that beyond $k_C$, the power spectrum can still grow slightly (say, with a much lower spectral index $p'\lesssim 2$), by a factor of a few, up to a maximum value at the scale which we may call $k_{PBH}$, corresponding to the scale which dominates PBH formation (see Fig. \ref{power}).
In this case, and assuming $P_\Phi(k_C) \sim 10^{-3}$, the effect on the orbital parameter of binaries is of order $\langle j^2 \rangle \sim (k_0/k_C)^2$, and becomes irrelevant unless $k_C \lesssim 10^3 k_0$. Since $k_0< 10^5 k_{eq}$ while 
$k_{PBH} \sim 10^9 k_{eq}$ for PBH in the stellar mass range, the effect of the steeply rising power spectrum on the eccentricity of binaries will only be important if this is crowned by a {\em broad} peak with a maximum at $k_{PBH} \gtrsim 10 k_C$. The corresponding effect on the merger rates is somewhat intricate in general, but it is only significant at low masses. For $m\lesssim 100$, the bound on the fraction of dark matter can relaxed to $f\lesssim 2\cdot 10^{-2}$ due to this effect, for sufficiently low $k_C\sim 10^{7} k_{eq}$. Interestingly, in this regime the merger rate becomes almost independent of $m$ (see Fig. \ref{caseC2}). This is in contrast with the standard situation where the angular momentum is supplied by a neighboring PBH, where we have $\Gamma \propto m^{-32/37}$. 

We have also investigated the dependence of merger rates on the constituent masses, with particular attention to the universality coefficient $\alpha$  \cite{Kocsis}. This is rather insensitive to the unknown initial PBH mass distribution function, and can be determined observationally with some precision (of order $15\%$ given a sufficiently large number of PBH merger events $\sim 10^3$ \cite{Kocsis,review}). It seems therefore of great empirical relevance for PBH scenarios. Our results for $\alpha$ are summarized in Fig. \ref{alpha}. In the case where cosmological perturbations are subdominant relative to the torque from neighboring PBH, it was argued in \cite{Kocsis} that $\alpha \approx 1\pm 0.05$ (this is displayed as a horizontal shading in Fig. \ref{alpha}). We point out, however, that this narrow range shifts to $\alpha \approx 1.26$ once the effect of PBH infall onto binaries is taken into consideration, by excluding disturbed binaries from the merger count (after the infall these binaries are likely to have a much larger lifetime than the present age of the universe).\footnote{The shift of $\alpha$ in the case where cosmological perturbations are subdominant is due to the factor $Y_{min}^{21/74}$ in Eq. (\ref{complete}). Note that the merger rates have an additional exponential dependence on the infall radius $Y_{min}$. Such dependence drops out from the universality coefficient $\alpha$ due to the linearity of $Y_{min}$ in the total mass $M$ [see Eq. (\ref{ymin})]. The assumption of linear behaviour seems very reasonable \cite{Ioka,RSVV}, but may require further validation from numerical simulations, since any departures from it may have a significant effect on $\alpha$. This study is outside the scope of the present work and is left for further research.}  On the other hand, 
in the case where cosmological perturbations dominate the variance of $j$, the coefficient $\alpha$ ranges from $15/16$ to $5/3$ depending on the value of the spectral index $p$ [see Fig. \ref{alpha}]. In this case, the value of $\alpha$ is unaffected by infalls.
We conclude that, as a matter of principle, an accurate measurement of merger rates of PBH might carry some information on the circumstances surrounding PBH binary formation, including the amplitude and spectral index of primordial perturbations on very short wavelengths or the effect of PBH infalls. 

In turn, any information on the primordial power spectrum which may be obtained through a measument of $\alpha$ would constrain the underlying inflationary dynamics, from the  scale of binaries down to the scale of PBH. This might complement other possible probes on the amplitude of $P_\Phi$ at small scales, such as upcoming searches for spectral distortions in the CMB, or pulsar timing array constraints \cite{PIXIE,current,BCP}. 

There are several directions in which our study could be extended. It was recently pointed out in \cite{KGB} that for $f\ll 1$ a dark matter dress may develop around PBH before they form binaries, and that this would affect the semi-major axis and the eccentricity of binaries after the first few oscilations which shake off most of the dark matter cloud. It was also shown that the effect on the merger rates is nonetheless small. In the present context, the effect may be even smaller, since in the presence of enhanced cosmological perturbations the merger rate is dominated by much smaller binaries forming deeper in the radiation era, when the halo around each PBH has had less time to accrete. Nonetheless, it might be interesting to be more quantitative about this effect, taking also into consideration the case of constituent masses which differ by a sizable factor, in order to assess the possible impact on $\alpha$. Finally, it was pointed out in \cite{RSVV} that for $f\gtrsim 0.1$ binaries can be disrupted by collision with compact N-body systems, which may substantially deplete the population of pristine binaries. Investigation of these issues seems to require further simulations, and remains an 
interesting direction of research.


\section*{Acknowledgements}

We thank V. Atal, N. Bellomo, J. Bernal, C. Byrnes, J. Miralda, M. Sasaki and T. Tanaka for useful discussions. This work has been partically supported by FPA2016-76005-C2-2-P, MDM-2014-0369 of ICCUB (Unidad de Excelencia Maria de Maeztu), and AGAUR2017-SGR-754. N.T. is supported by an INPhINIT grant 
from Òla CaixaÓ Foundation (ID 100010434) code LCF/BQ/IN17/11620034. This project has also received funding from the
European UnionÕs Horizon 2020 research and innovation programme under the Marie Sklodowska-Curie grant agreement No. 713673.
We have used the publicly available WebPlotDigitizer \cite{wpd} to digitise plots.

\end{document}